\documentclass[10pt,a4paper]{article}

\usepackage{epsfig,graphicx}
\usepackage{amsmath,amsfonts,amssymb}
\usepackage{citesort}

\newcommand{\unit}{\leavevmode\hbox{\small1\kern-3.6pt\normalsize1}}

\def\lsim{\raise0.3ex\hbox{$\;<$\kern-0.75em\raise-1.1ex\hbox{$\sim\;$}}}
\def\gsim{\raise0.3ex\hbox{$\;>$\kern-0.75em\raise-1.1ex\hbox{$\sim\;$}}}
\newcommand{\wh}{\widehat}

\newcommand{\vev}[1]{\langle #1\rangle}

\begin{document}

\begin{flushright}
FTUAM 05/18\\
IFT-UAM/CSIC-05-49\\
\vspace*{3mm}
{\today}
\end{flushright}

\vspace*{5mm}
\begin{center}
{\Large \textbf{Phenomenological viability of orbifold models\\ with
    three Higgs families} }

\vspace{0.5cm} 
{\large N.~Escudero, C.~Mu\~noz and A.~M.~Teixeira}\\[0.2cm]
{\textit{Departamento de F\'{\i}sica Te\'{o}rica C-XI, \\
 Universidad Aut\'{o}noma de Madrid,
Cantoblanco, E-28049 Madrid, Spain}}\\

\vspace*{0.2cm} 
{\textit{Instituto de F\'{\i}sica Te\'{o}rica C-XVI, \\
Universidad Aut\'{o}noma de Madrid,
Cantoblanco, E-28049 Madrid, Spain}}\\[0pt]

\vspace*{0.3cm} 
\begin{abstract}
We discuss the phenomenological viability of string 
multi-Higgs doublet models, namely a scenario of heterotic $Z_3$ 
orbifolds with two Wilson lines, which naturally predicts three 
supersymmetric families of matter and Higgs fields.
We study the orbifold parameter space, and discuss the compatibility of the
predicted Yukawa couplings with current experimental data.
We address the implications of tree-level flavour changing neutral processes
in constraining the Higgs sector of the model, finding that viable scenarios
can be obtained for a reasonably light Higgs spectrum.
We also take into account the tree-level contributions to indirect CP
violation, showing that the experimental value of $\varepsilon_K$ can
be accommodated in the present framework. 
\end{abstract}
\end{center}
 

\section{Introduction}\label{intro}

The understanding of the observed pattern of quark and lepton masses
and mixings remains as one of the most important open questions in
particle physics. 
From experiment, we believe that Nature contains three
families of quarks and leptons, with peculiar mass hierarchies. 
Moreover, there is firm evidence that the flavour structure
in both quark and lepton sectors is far from trivial, as exhibited by
the current bounds on the quark~\cite{pdg2004} and
lepton~\cite{Maltoni:2004ei,Strumia:2005tc,Fogli:2005cq}  
mixing matrices.

The standard model of strong and electroweak interactions (SM) fails in
explaining some important issues such as the fermion flavour
structure or the number of fermion families we encounter in Nature.
Moreover, the mechanism of mass generation for quarks and
leptons is still unconfirmed, 
since the Higgs boson is yet to be discovered in a collider.

Other high-energy motivated theories, such as supersymmetry (SUSY),
supergravity (SUGRA), or grand unified theories (GUTs) may repair some 
shortcomings
of the SM (as for instance the hierarchy problem, the existence of
particles with distinct spin, or the unification of gauge interactions), 
but they still fail in providing a clear understanding
of the nature of masses, mixings and number of families.
In this sense, a crucial ingredient to relate theory and observation
is the precise knowledge of how fermions and Higgs scalars interact,
in other words, the Yukawa couplings of the fundamental theory.

String theory is the only candidate to unify all known interactions
(strong, electroweak and gravitational) in a consistent way, 
and therefore it must necessarily contain the SM as its low-energy
limit. In this sense, string theory must provide an answer to the
above mentioned questions. 
A very interesting method to obtain a
four dimensional effective theory is the compactification of the $E_8
\times E_8$ heterotic string~\cite{Gross:1984dd} on six-dimensional
orbifolds~\cite{Dixon:1986jc}, and this has proved to be a very successful
attempt at finding the superstring standard 
model~\cite{Ibanez:1986tp,Ibanez:1987sn,Bailin:1987xm,Ibanez:1987pj,Casas:1987us,Font:1988tp,Kim:1988dd,Casas:1988se,Casas:1988hb,Font:1988mm,Casas:1988vk,Casas:1988wy,Font:1989aj,Casas:1989wu,Katsuki:1989bf,Kim:1992en,Aldazabal:1995cf,Munoz:2001yj,Abel:2002ih,Kobayashi:2004ya,Kobayashi:2005vb,Buchmuller:2005jr}
(other interesting attempts at model building using Calabi-Yau
spaces~\cite{Candelas:1985en}, fermionic constructions~\cite{Kawai:1986va,Antoniadis:1986rn}, and heterotic
M-theory ~\cite{Horava:1995qa,Witten:1996mz}, can be found in Refs.~\cite{Greene:1986ar,Donagi:2004ub,Braun:2005ux},~\cite{Antoniadis:1987tv,Faraggi:1989ka,Cleaver:1998sa,Cleaver:2001ab,Chaudhuri:1994cd,Chaudhuri:1995ve},
and~\cite{Donagi:1999ez}, respectively).
As it was shown in~\cite{Ibanez:1987sn,Ibanez:1987pj}, the
use of two Wilson
lines~\cite{Dixon:1986jc,Ibanez:1986tp} on the torus defining
a symmetric $Z_3$ orbifold can give rise to SUSY models with
$SU(3)\times SU(2)\times U(1)^n$ gauge group and three families of
chiral particles with the correct $SU(3)\times SU(2)$ quantum numbers.
These models present very attractive features from a phenomenological
point of view. 
One of the $U(1)$s of the extended gauge group is in general
anomalous, and it can induce a Fayet-Iliopoulos (FI) 
$D$-term~\cite{Witten:1984dg,Dine:1987xk,Atick:1987gy,Dine:1987gj} that
would break SUSY at very high energies (FI scale $\sim
\mathcal{O}(10^{16-17}$ GeV)). To preserve SUSY, some fields
will develop a vacuum 
expectation value (VEV) to cancel the undesirable $D$-term. The FI
mechanism allows to break the
gauge group down to $SU(3)_c\times SU(2)_L\times U(1)_Y$ and obtain the
mass spectrum of the minimal supersymmetric standard model (MSSM),
plus some exotic matter, as extra singlets, doublets or vector-like
triplets, depending on the model, as shown in 
Refs.~\cite{Casas:1988hb,Font:1988mm} and~\cite{Casas:1987us}.

Orbifold compactifications have other remarkable properties.
For instance, they provide a geo\-me\-tric mechanism to
generate the mass hierarchy for quarks and 
leptons~\cite{Hamidi:1986vh,Dixon:1986qv,Ibanez:1986ka,Casas:1989qx,Casas:1992zt}
through renormalisable Yukawa couplings. 
$Z_n$ orbifolds have twisted fields which are attached to the 
orbifold fixed points. Fields at different fixed points may
communicate with each other only by world sheet instantons. The
resulting renormalisable Yukawa couplings can be explicitly 
computed~\cite{Hamidi:1986vh,Dixon:1986qv,Casas:1990hi,Burwick:1990tu,Kobayashi:1991rp,Casas:1991ac}
and they receive exponential suppression factors that depend on the 
distance between the fixed points to which the relevant fields are 
attached. These distances can be varied by giving different VEVs
to the $T$-moduli associated with the size 
and the shape of the orbifold. 

However, the major 
problem that one encounters when trying to obtain models with entirely 
renormalisable Yukawas lies at the phenomenological level, and is
deeply related to obtaining the correct quark mixing. 
Summarising the analyses of Refs.~\cite{Casas:1989qx,Casas:1992zt}, for
prime orbifolds the space group selection rules and the need for a 
fermion hierarchy
forces the fermion mass matrices to be diagonal at the renormalisable
level. Thus, in these cases, the Cabibbo-Kobayashi-Maskawa (CKM)
parameters must arise at the non-renormalisable level.
For analyses of non-prime orbifolds see 
Refs.~\cite{Casas:1989qx,Casas:1992zt,Ko:2004ic,Ko:2005sh}.

For example, since the FI breaking generates 
VEVs  for fields of order $\langle\chi_j \rangle\sim 10^{16-17}$ GeV, if
one has terms in the superpotential
of the type $\frac{1}{M_P^{m}}\,\chi_1\cdot\cdot\cdot\chi_m\,\xi\,\xi\,\xi$, 
these would produce couplings of order $(M_{FI}/M_P)^{m} $.
Therefore, depending on $m$, different values for the couplings
might be generated. Obviously, the presence of these couplings
is very model-dependent and introduces a high degree of uncertainty
in the computation. However, it is important to remark that
having the latter couplings
is not always allowed in string constructions. 
First of all, they must be gauge-invariant, something that is not
easy to achieve, due to the large number of $U(1)$ charges which are 
associated to the particles in these models.
Even if the couplings fulfil this condition, this does not mean that
they are automatically allowed. They must still fulfil   
the so-called ``stringy'' selection rules.
For example in the 
$SU(3)\times SU(2)\times U(1)_Y\times SO(10)_{hidden}$ model
of Ref.~\cite{Casas:1988hb}, where 
renormalisable couplings are present, only a small number 
of non-renormalisable terms are allowed by gauge invariance.
Nevertheless, even the latter terms turn out to be forbidden by 
string selection rules.

Clearly, purely renormalisable Yukawa couplings are preferable,
because, in general, due to the
arbitrariness of the VEVs of the fields entering the
non-renormalisable couplings, the predictivity is lost. 
Furthermore, as discussed above,
higher-order operators such as those induced by the FI 
breaking are 
very model-dependent. 
One possibility of avoiding the necessity of 
these non-renormalisable couplings is to relax 
the requirement of a minimal matter content (with
just two Higgs doublets) in a $Z_3$ orbifold with two Wilson lines.
Since these models naturally contain 
three families of everything, including Higgses, additional Yukawa
couplings will be present, with the possibility of leading to
realistic fermion masses and mixings, entirely at the renormalisable
level (with a key role being played by the FI breaking)~\cite{Abel:2002ih}.  
In addition, and
given the existence of three families of quarks and leptons, having
also three families of Higgses renders these models very aesthetic. In
fact, let us recall that experimental data imposes no constraints on the
number of Higgs families. 
Moreover, this non-minimal Higgs content, provided that the extra doublets are
light enough to be present at low-energies, also favours the
unification of gauge couplings in heterotic string constructions.
Due to the FI scale, the gauge couplings may unify at the string scale
($\approx g_{\mathrm{GUT}} \times 5.27\, 10^{17}$ GeV)~\cite{Munoz:2001yj}.

Thus, this class of string compactifications is one of the scenarios
where one can obtain a SM/MSSM compatible low-energy theory, albeit
with an extended Higgs sector. Furthermore it offers a solution to the
flavour problem of the SM and MSSM, since the structure of the Yukawa
couplings is completely derived from the geometry of the high-energy
string construction.
Given the increasing experimental
accuracy, accommodating the data on quark masses and mixings is not
straightforward. In this work, we propose to investigate in detail whether or
not it is possible to obtain $Z_3$ orbifold configurations that successfully
reproduce the observed flavour pattern in Nature. 
In this sense, having additional Yukawa couplings presents several
advantages, as for example
a greater flexibility when fitting the data from the quark masses and
mixings. 

On the other hand, when working in a multi-Higgs context, we should also
take into account the potential appearance of flavour-changing neutral currents
(FCNCs) at the tree level, which could contribute to a wide variety of
Higgs decays and interactions with other 
particles~\cite{Georgi:1978ri,McWilliams:1980kj,Shanker:1981mj,Flores:1982pr,Cheng:1987rs,Ellis:1986ip,Drees:1988fc,Griest:1989ew,Griest:1990vh,Haber:1989xc,Sher:1991km,Krasnikov:1992gd,Antaramian:1992ya,Nelson:1993vc,Masip:1995sm,Masip:1995bq,Aranda:2000zf,Escudero:2005hk}.
Generally, the most stringent limit to these flavour-changing  
processes is assumed to come from the mass difference of the long- 
and short-lived 
neutral kaons, $\Delta m_K=m_{K_L}-m_{K_S}$. A possible way to
overcome this problem is by imposing that the Higgs spectrum is heavy
enough to suppress the undesired contributions to the neutral meson
mass differences. 
As we will see, the Yukawa couplings of this $Z_3$ scenario
exhibit a strongly hierarchical structure, and this property is
instrumental in circumventing the FCNC problem without the need for an
excessively heavy Higgs sector.

$Z_3$ orbifolds are also very attractive when addressing
the lepton sector, and in fact offer an appealing scenario to study the
problem of neutrino masses (predicting naturally small Dirac masses, 
and thus a low see-saw scale). We postpone this analysis to a forthcoming 
work~\cite{EJMT}. 

This work is organised as follows. In Section~\ref{yukawa}, we
describe the main properties of the Yukawa couplings in $Z_3$ orbifold
models. We study the relations between the several orbifold parameters
induced from the quark mass hierarchy and from electroweak symmetry
breaking. Section~\ref{higgsphenom} is devoted to a brief overview of
the extended Higgs sector. In Section~\ref{yukint} we present the
contributions of neutral Higgs exchange to tree-level
FCNCs. The numerical analyses of the orbifold parameter space and
FCNCs in association with specific Higgs textures is given in 
Section~\ref{results}, where we also address the possibility of new 
contributions to indirect CP violation. Finally, we summarise our results in
Section~\ref{conc}.

\section{Yukawa couplings in $\pmb{Z_3}$ orbifold models}\label{yukawa}
In this Section we review some of the most relevant features of the
geometrical construction of the $Z_3$ orbifold 
leading to the computation of the quark mass
matrices. We study the correlations of the orbifold parameters arising
from the quark mass hierarchy and electroweak (EW) symmetry breaking
conditions, and derive useful relations which play an important role
in constraining the parameter space. 

\subsection{$\pmb{Z_3}$ orbifold: a brief review}\label{Z3}  

\begin{figure}[t]
  \begin{center} 
	\psfig{file=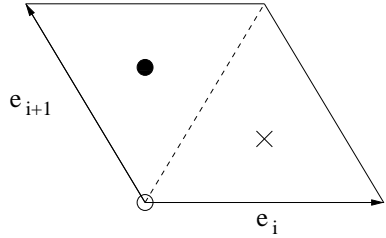,width=50mm,angle=0,clip=} 
    \caption{Two dimensional sublattices ($i=1,3,5$) of the $Z_3$
    orbifold, symbolically denoting the fixed points as
    ($\circ,\bullet,\times$).}\label{fig:orbipicture}
\end{center}
\end{figure}

The $Z_3$ orbifold is constructed by dividing $R^6$ by the $[SU(3)]^3$
root lattice modded by the point group (P) with generator $\theta$,
where the action of $\theta$ on the lattice basis is 
$\theta e_i = e_{i+1}$, $\theta e_{i+1} = -(e_i + e_{i+1})$, with
$i=1,3,5$. The two-dimensional sublattices associated to $[SU(3)]^3$
are presented in Fig.~\ref{fig:orbipicture}. 
Invariance under the point group reduces the orbifold deformation 
parameters to nine: the three radii of the sublattices and the 
six angles between
complex planes. The latter parameters correspond to the
VEVs of nine singlet fields appearing in the spectrum of the untwisted
sector, and which have perturbatively flat potentials. These so-called
moduli fields are usually denoted by $T$.

In orbifold constructions, twisted strings appear attached to fixed
points under the point group. In the case of the $Z_3$ orbifold there
are 27 fixed points under P, and therefore 27 twisted
sectors. We will denote the three fixed points of each two-dimensional
lattice as in Fig.~\ref{fig:orbipicture}. 
In the $Z_3$ orbifold, the general form of the Yukawa couplings 
between the twisted fields is given by a Jacobi theta function, and
their expressions can be found, for example, in the Appendix of
Ref.~\cite{Casas:1991ac}. 
The Yukawa couplings contain suppression factors that depend
on the relative positions of the fixed points to which the the fields
involved in the coupling are attached, and on the size and shape of
the orbifold (i.e. the deformation parameters). Let us first study the
situation before taking into account the effect of the FI
breaking. Let us suppose that the two non-vanishing Wilson lines
correspond to the first and second sublattices. Then, the 27 twisted
sectors come in nine sets with three equivalent sectors in each
one. The three generations of matter (including Higgses) correspond to
changing the third sublattice component ($\circ,\bullet,\times$) of
the fixed point, while keeping the other two fixed. Consider for
example the following assignments of observable matter to fixed point
components in the first two sublattices,
\begin{align}
&Q \, \leftrightarrow \, \circ \,\circ\,
\quad \quad 
u^c \, \leftrightarrow \, \circ\,\circ\,
\quad \quad 
d^c \, \leftrightarrow \, \times\,\circ\,
\nonumber \\
&L \, \leftrightarrow \, \bullet\,\bullet\,
\quad \quad 
e^c \, \leftrightarrow \, \bullet\,\times\,
\quad \quad 
\nu^c \, \leftrightarrow \, \times\,\times\,
\nonumber \\
& \quad \quad \quad H^u \, \leftrightarrow \, \circ\,\circ\,
\quad \quad 
H^d \, \leftrightarrow \, \bullet\,\circ\,\quad \quad \quad \,.
\end{align}
In this case, the up- and
down-quark mass matrices are given by~\cite{Abel:2002ih}:
\begin{equation}\label{Z3mass:beforeFY}
\mathcal{M}^u= 
g \,N \,A^u\,, \quad \quad
\mathcal{M}^{d}= 
g \,N \,\varepsilon_1 \,A^d\,,
\end{equation}
where
\begin{equation}\label{AuAd:bfFI}
A^u = \left(
\begin{array}{ccc}
w_2 & w_6 \,\varepsilon_5 & w_4 \,\varepsilon_5 \\
w_6 \,\varepsilon_5 & w_4 & w_2 \,\varepsilon_5 \\
w_4 \,\varepsilon_5 & w_2 \,\varepsilon_5 & w_6
\end{array}\right)\, , \quad \quad
A^d = \left(
\begin{array}{ccc}
w_1 & w_5 \,\varepsilon_5 & w_3 \,\varepsilon_5 \\
w_5 \,\varepsilon_5 & w_3 & w_1 \,\varepsilon_5 \\
w_3 \,\varepsilon_5 & w_1 \,\varepsilon_5 & w_5
\end{array}\right)\,,
\end{equation}
$g$ is the gauge coupling constant, and $N$ is related to the 
volume of the $Z_3$ lattice unit cell such that $g\,N \approx 1$. 
In the above matrices 
$w_i$ denote the VEVs of the neutral components of the six Higgs 
doublet fields\footnote{For convenience, we adopt a distinct notation
from the one originally used in Ref.~\cite{Abel:2002ih}. Instead of
denoting the Higgs fields (and the corresponding VEVs) 
$H^{d,u}_i\, (v_i^{d,u})$, $i=1,2,3$, we use 
$H_i\, (w_i)$, $i=1,\ldots ,6$, with odd (even) cases referring to down-
(up-) quark coupling Higgses.}. Since we are assuming an orthogonal
lattice, i.e. with the six angles equal to zero, only the 
diagonal moduli ($T_i$), which are related with the radii of 
the three sublattices\footnote{
Actually, in Ref.~\cite{Abel:2002ih} it is argued that it is possible
to fit the entire fermion data by using only one degenerate radius. In
our approach, and given the increasing accuracy in the experimental
data, we allow for the more general case of three radii (and thus
three moduli).\label{Tifoot}}, 
contribute to the Yukawa couplings, through the
suppression factors $\varepsilon_i$
\begin{equation}\label{eiTi}
\varepsilon_i \,\approx\,
3 \,e^{-\frac{2 \pi}{3} T_i}\,\quad \quad i,j=1,3,5\,. 
\end{equation}

\subsection{Quark mass matrices and Yukawa couplings after the 
Fayet-Iliopoulos breaking}\label{yukawa:matrices}  
As mentioned in the Introduction, the anomalous $U(1)$ of the extended 
$SU(3) \times SU(2) \times U(1)^n$ gauge group generates a
Fayet-Iliopoulos $D$-term which could in principle break SUSY at
energies close to the string scale.
This term can be cancelled when 
scalar fields ($C_i$), which are singlets under $SU(3) \times SU(2)$,
develop large VEVs ($10^{16-17}$ GeV). The VEVs of these fields
($c_i$), have several important effects. Firstly, they 
break the original 
$SU(3) \times SU(2) \times U(1)^n$ gauge group down to the (MS)SM 
$SU(3) \times SU(2) \times U(1)$. Secondly, they induce very large
effective mass terms for many particles (vector-like triplets and doublets, as
well as singlets), which thus decouple from the low-energy
theory. Even so, the SM matter remains massless, surviving as the 
zero mass mode 
of combinations with the other (massive) states.
All these effects modify 
the mass matrices of the low-energy effective theory
(see Eq.~(\ref{Z3mass:beforeFY})), which, for the example studied
in~\cite{Abel:2002ih}, are now given by\footnote{Note that although
  the $c_i$ are in general complex VEVs, they only introduce a global
  and therefore unphysical phase in the mass matrix. More complicated
  examples would in principle give rise to a contribution to the CP
  phase~\cite{Abel:2002ih}. This mechanism to generate the CP phase
  through the VEVs of the fields cancelling the FI $D$-term was used
  first, in the context of non-renormalisable couplings, in
  Ref.~\cite{Casas:1989qx}. For a recent analysis, see
  Ref.~\cite{Giedt:2000es}.\label{CPfoot}}
\begin{align}\label{quark:mass}
\mathcal{M}^u=& \,
g \,N \,a^{u^c} \,A^u \,B^{u^c}\,, \nonumber \\
\mathcal{M}^d=& \,
g \,N \varepsilon_1 \,a^{d^c}\, A^d \,B^{d^c}\,,
\end{align}
where $A^{u,d}$
are the quark mass matrices prior to FI breaking (see
Eq.~(\ref{AuAd:bfFI})), $a^{f}$ is given by 
\begin{equation}\label{af:def}
a^{f}\,=\,\frac{\hat c^f_2}{\sqrt{|\hat c_1^f|^2+|\hat c_2^f|^2}}\,,
\end{equation}
with $f=u^c,d^c$, and $B^f$ is the diagonal matrix defined as 
\begin{equation}\label{quark:B}
B^f\,=\, \operatorname{diag}\, (\,\beta^f \,\varepsilon_5,\, 1\,, 
\alpha^f/\varepsilon_5\,)\,. 
\end{equation}
Finally 
\begin{align}\label{alpha:beta:def}
\alpha^f = \varepsilon_5 \sqrt{\frac{|\hat c_1^f|^2+|\hat c_2^f|^2}{
|\hat c_1^f \varepsilon_5|^2+|\hat c_2^f|^2}}\,, \quad \quad
\beta^f = \sqrt{\frac{|\hat c_1^f|^2+|\hat c_2^f|^2}{
|\hat c_1^f|^2+|\hat c_2^f \varepsilon_5|^2}}\,.
\end{align}
In the above, $\hat c_i^f$ are derived from the VEVs of the heavy fields
responsible for the FI breaking as 
\begin{equation}\label{hatc:def}
\hat c_1^f \,\equiv \,\varepsilon^{\prime (f)}\, c_1^f\,,\quad \quad 
\hat c_2^f \,\equiv \,\varepsilon^{\prime \prime (f)}\,c_2^f\,,
\end{equation}
where in each case $ \varepsilon^{\prime}$ and 
$ \varepsilon^{\prime \prime}$ can take any of the following values:
\begin{equation}\label{eprime:def}
\varepsilon^{\prime}\,, \,\, \varepsilon^{\prime \prime}
\equiv
1, \,\varepsilon_1, \,\varepsilon_3, \,\varepsilon_1 \,\varepsilon_3\,.
\end{equation}
Let us also stress that one should not take $\alpha^f$, $\beta^f$,
$\varepsilon_5$ and $a^f$ as independent parameters. 
In fact, Eqs.~(\ref{af:def},\ref{alpha:beta:def}) imply that 
\begin{equation}\label{af:alpha:beta}
a^f\,=\, 
\frac{\left( 1-{\alpha^f}^2\right)^{1/2}}{\alpha^f} \, 
\frac{\varepsilon_5}{\left( 1-\varepsilon_5^2 \right)^{1/2}}
\,=\,
\left( 1-\frac{1}{{\beta^f}^2}\right)^{1/2} \,
\frac{1}{\left( 1-\varepsilon_5^2 \right)^{1/2}}\,,
\end{equation}
so that for given values of $\varepsilon_5$ and $\alpha^f$, $\beta^f$
is fixed as
\begin{equation}\label{bf:alpha}
\beta^f\,=\,\frac{1}{\sqrt{1+\varepsilon_5^2
 \left(1-\frac{1}{{\alpha^f}^2} \right)}}\,.
\end{equation}
Eqs.~(\ref{quark:mass},\ref{quark:B}) become more transparent when
the terms that encode the flavour structure are explicitly displayed:
\begin{equation}\label{quark:aFI}
A^u B^{u^c}= \left(
\begin{array}{ccc}
w_2\,\varepsilon_5 \, \beta^{u^c}& w_6 \,\varepsilon_5 & w_4 \, \alpha^{u^c} \\
w_6 \,\varepsilon_5^2 \, \beta^{u^c}& w_4 & w_2\, \alpha^{u^c} \, \\
w_4 \,\varepsilon_5^2 \, \beta^{u^c}& w_2\,\varepsilon_5 
& w_6\, \alpha^{u^c}/\varepsilon_5
\end{array}\right), \,\,\,
A^d B^{d^c}= \left(
\begin{array}{ccc}
w_1\,\varepsilon_5 \, \beta^{d^c}& w_5 \,\varepsilon_5 & w_3 \, \alpha^{d^c} \\
w_5 \,\varepsilon_5^2 \, \beta^{d^c}& w_3 & w_1\, \alpha^{d^c} \, \\
w_3 \,\varepsilon_5^2 \, \beta^{d^c}& w_1\,\varepsilon_5 
& w_5\, \alpha^{d^c}/\varepsilon_5
\end{array}\right).
\end{equation}
Given that the mass matrices are related to the Yukawa couplings as 
\begin{equation}\label{quark:yuk:mass}
\mathcal{M}^u = \sum_{i=2,4,6} w_i \,Y^u_i\,, \quad \quad 
\mathcal{M}^d = \sum_{i=1,3,5} w_i \,Y^d_i\,, 
\end{equation}
the structure of the Yukawa couplings is easily derived from
Eq.~(\ref{quark:aFI}). 
For the down sector, the latter read:
\begin{align}\label{Yd:1:5}
Y^d_1 = g N \varepsilon_1 a^{d^c} \,
\left(\begin{array}{ccc}
\varepsilon_5 \, \beta^{d^c}& 0&0 \\
0 & 0 & \alpha^{d^c} \\
0& \varepsilon_5 & 0 
\end{array}
\right)\,,\quad 
 Y^d_3 = g N \varepsilon_1 a^{d^c} \,
 \left(\begin{array}{ccc}
 0 & 0 & \alpha^{d^c} \\
 0 & 1 & 0\\
 \varepsilon_5^2 \, \beta^{d^c}& 0&0
 \end{array}
 \right)\,,\nonumber 
\end{align}
\begin{align}
  Y^d_5 = g N \varepsilon_1 a^{d^c} \,
   \left(\begin{array}{ccc}
   0 &  \varepsilon_5 & 0 \\
   \varepsilon_5^2 \, \beta^{d^c}& 0 & 0\\
   0 & 0 & \alpha^{d^c}/\varepsilon_5
   \end{array}
  \right)\,.
\end{align}
The Yukawa couplings for the up-type quarks can be also obtained
by doing the appropriate replacements: $(\varepsilon_1 a^{d^c}) \to
a^{u^c}$ and $\alpha^{d^c}, \beta^{d^c} \to \alpha^{u^c}, \beta^{u^c}$.

Expanding the eigenvalues of the quark mass matrices up to leading
order in  $\varepsilon_5$, one can derive the following
relation\footnote{
Regarding quark mixing, it is also possible to obtain analytical
expressions (up to second order in $\varepsilon_5$) for the several CKM
matrix elements, as done in Ref.~\cite{Abel:2002ih}.} for the Higgs
VEVs in terms of the quark masses\footnote{Notice that there
is a misprint in these equations in Ref.~\cite{Abel:2002ih}, where in
the corresponding version of Eq.~(\ref{vev:quarkmass}) the factor
$\varepsilon_5^5$ appeared as $\varepsilon_5^2$.}~\cite{Abel:2002ih} 
\begin{align}\label{vev:quarkmass}
\text{down-quarks\,:}\,\,\,
&\{w_1,w_3,w_5\} \,(g N \,\varepsilon_1 \,a^{d^c})\,=
\left\{
\frac{1}{\varepsilon_5 \beta^{d^c}} \left(m_d + \varepsilon_5^5
\frac{m_b^2}{m_s}\right), m_s, \frac{m_b \varepsilon_5}{\alpha^{d^c}}
\right\}\,, \nonumber\\
\text{up-quarks\,:}\,\,\,
&\{w_2,w_4,w_6\} \, (g N \,a^{u^c})\,=
\left\{
\frac{1}{\varepsilon_5 \beta^{u^c}} \left(m_u + \varepsilon_5^5
\frac{m_t^2}{m_c}\right), m_c, \frac{m_t \varepsilon_5}{\alpha^{u^c}}
\right\}\,.
\end{align}

The most striking effect of the FI
breaking is that it enables the reconciliation of the Yukawa couplings
predicted by this scenario with experiment. 
In particular, and as we will see in Section~\ref{quark:orbifold}, 
the quark spectra and a successful CKM matrix can now be accommodated. 

\subsection{EW symmetry breaking and the orbifold parameter
  space}\label{ewsb:orbifold} 
In addition to the hierarchy constraint imposed by the observed pattern
of quark masses, the VEVs must further comply with other constraints
as those arising from EW symmetry breaking (EWSB):
\begin{equation}\label{ewz}
w_1^2+w_2^2+w_3^2+w_4^2+w_5^2+w_6^2=2\,M_Z^2/(g^2+g'^2)
\approx (174\text{ GeV})^2\,.
\end{equation}
In particular, we have that
\begin{align}\label{wmz}
&\frac{1}{(g N a^{u^c})^2}\,\left[
\frac{1}{(\varepsilon_5 \beta^{u^c})^2} \left(m_u + \varepsilon_5^5
\frac{m_t^2}{m_c}\right)^2+
m_c^2 +  \left(\frac{m_t \varepsilon_5}{\alpha^{u^c}}\right)^2
\right]+
\nonumber\\
&\frac{1}{(g N \varepsilon_1 a^{d^c})^2}\,\left[
\frac{1}{(\varepsilon_5 \beta^{d^c})^2} \left(m_d + \varepsilon_5^5
\frac{m_b^2}{m_s}\right)^2+
m_s^2 +  \left(\frac{m_b \varepsilon_5}{\alpha^{d^c}}\right)^2
\right] \approx (174 \, \text{GeV})^2\,.
\end{align}
We notice that the above condition can always be fulfilled since the
quark Yukawa matrix prefactors, $\varepsilon_1$ and $gN$, 
have not yet been used. 
At this point, let us introduce a generalised definition for $\tan
\beta$:
\begin{equation}\label{tb}
\tan \beta \,=\, \frac{v_u}{v_d}\,\equiv\, 
\frac{\sqrt{w_2^2+w_4^2+w_6^2}}{\sqrt{w_1^2+w_3^2+w_5^2}}\,.
\end{equation}
Using Eq.~(\ref{vev:quarkmass}), Eq.~(\ref{tb}) can be rewritten as
\begin{align}\label{wmbeta}
\tan \beta \,=\, 
\varepsilon_1 \frac{a^{d^c}}{a^{u^c}} \,\sqrt{
\frac{
\frac{1}{(\varepsilon_5 \beta^{u^c})^2} \left(m_u + \varepsilon_5^5
\frac{m_t^2}{m_c}\right)^2+
m_c^2 +  \left(\frac{m_t \varepsilon_5}{\alpha^{u^c}}\right)^2
}{
\frac{1}{(\varepsilon_5 \beta^{d^c})^2} \left(m_d + \varepsilon_5^5
\frac{m_b^2}{m_s}\right)^2+
m_s^2 +  \left(\frac{m_b \varepsilon_5}{\alpha^{d^c}}\right)^2
}}\,.
\end{align}
From the above equation it becomes manifest that by considering a
given value for $\tan \beta$ we are implicitly defining
$\varepsilon_1$, for fixed values of $\varepsilon_5$ and $\alpha^f$.
This in turn implies that according to Eq.~(\ref{wmz}), 
$g\,N$ is in fact 
a function of $\tan \beta$, $\varepsilon_5$ and $\alpha^f$, and its
value, $g\,N \approx 1$, suffers tiny fluctuations (of order 1\% - 10\%) in
order to accommodate the correct EWSB.
The latter statements become more transparent noticing that by 
bringing together Eqs.~(\ref{wmz})~and~(\ref{wmbeta}), one can derive useful
relations that allow to express $gN$ and $\varepsilon_1$
as a function of the quark masses and orbifold parameters for a given
value\footnote{Whenever referring to a parameter whose
  value was estimated using the EWSB conditions, and which is a
  function of $\tan \beta$, we will use the designation ``EWSB
  fit''.} of $\tan \beta$:
\begin{align}\label{auad:rel}
gN &= \frac{1}{a^{u^c}}\,
\frac{\left( 1+\tan^2 \beta \right)^{1/2}}{\tan \beta}\,\,
\frac{\sqrt{\frac{1}{(\varepsilon_5 \beta^{u^c})^2} \left(m_u + \varepsilon_5^5
\frac{m_t^2}{m_c}\right)^2+
m_c^2 +  \left(\frac{m_t \varepsilon_5}{\alpha^{u^c}}\right)^2
}}{174\, \text{GeV}}\,, \nonumber \\
& \nonumber \\
\varepsilon_1 gN &=\frac{1}{a^{d^c}}\,
{\left( 1+\tan^2 \beta \right)^{1/2}}\,\,
\frac{\sqrt{\frac{1}{(\varepsilon_5 \beta^{d^c})^2} \left(m_d + \varepsilon_5^5
\frac{m_b^2}{m_s}\right)^2+
m_s^2 +  \left(\frac{m_b \varepsilon_5}{\alpha^{d^c}}\right)^2
}}{174\, \text{GeV}}\,.
\end{align}
The first equality of Eq.~(\ref{auad:rel}) 
provides a clear insight to understanding the smallness of the
fluctuations of $g\,N$. Assuming the limit where $\alpha^{u^c},
\varepsilon_5 \ll 1$, $a^{u^c} \sim \varepsilon_5/\alpha^{u^c}$, so
  that $g\,N \approx m_t /(174$ GeV).

It is also important to
comment on the relative size of the VEVs $\hat c_1$
and $\hat c_2$. From the definition of $a^f$ (Eq.~(\ref{af:def})) we
can derive an additional relation
\begin{equation}
|c_1^f|\,=\,
\frac{\varepsilon^{\prime \prime (f)}}{\varepsilon^{\prime (f)}}\,
\sqrt{\frac{1-{a^f}^2}{{a^f}^2}}\, |c_2^f|\,,
\end{equation}
where we have used the definitions of
Eqs.~(\ref{hatc:def},\ref{eprime:def}).
If, for example, one assumes the VEVs to be of the same order of
magnitude, i.e. $c_1 \sim c_2$, then one should further ensure that 
\begin{equation}
\frac{\varepsilon^{\prime \prime (f)}}{\varepsilon^{\prime (f)}}\,
\sqrt{\frac{1-{a^f}^2}{{a^f}^2}}\, \sim 1\,.
\end{equation}

\vspace*{3mm}
To conclude this Section, 
let us make a few remarks regarding two topics so far not discussed.
Firstly, and since it is well known that the CP symmetry is not
conserved in nature, it is important to comment on the sources of CP violation
present in this class of models. The Yukawa couplings have been defined
through real quantities, so that no physical phase appears via the CKM
mechanism. However, this need not be the most general scenario.
Dismissing for the present time the possibility of
spontaneous CP violation, associated with non-trivial phases of the Higgs VEVs,
there still remains another source of CP violation, in addition to the
one already mentioned in footnote~\ref{CPfoot}. 
Should the VEV of the moduli field have a phase, then CP (which is a
gauge symmetry of the model) would be spontaneously broken at very high
energies. The phases would be fed into $\varepsilon_i$ (thus also 
appearing in $\alpha^f$), and would be present in the Yukawa couplings.
Therefore, in the low-energy theory, CP would be explicitly violated
via the usual CKM mechanism~\cite{Acharya:1995ag,Bailin:1998xx}.

It is also relevant to mention the effect of the
renormalisation group equations (RGE) on the mass matrices presented in this
Section. The flavour structure of Eqs.~(\ref{quark:mass},\ref{quark:aFI}) 
is associated with a mechanism taking place at a very
high energy scale. However, and given the clearly hierarchical
structure of the quark mass matrices, one does not expect that 
RGE running will significantly affect the predictions of the model.

\section{The extended Higgs sector}\label{higgsphenom}
As mentioned in the previous sections, in this class of orbifold models,
one has replication of families in the Higgs sector.  
By construction, this scenario contains three generations of
$SU(2)$ Higgs doublet superfields, with hypercharge $-1/2$ and $+1/2$, 
respectively coupling to down- and up- type quarks.
\begin{equation}\label{H:superf}
\widehat{H}_{1(3,5)}=
\left( \begin{array}{c}
\widehat{h}^0_{1(3,5)} \\
\widehat{h}^-_{1(3,5)}
\end{array} \right)\,,  \quad \quad
\widehat{H}_{2(4,6)}=
\left( \begin{array}{c}
\widehat{h}^+_{2(4,6)} \\
\widehat{h}^0_{2(4,6)}
\end{array} \right)\,.
\end{equation}
In Ref.~\cite{Escudero:2005hk}, we have studied the general case of SUSY
models with Higgs family replication, as is the present case.
Hence, we will just summarise here some important features which are
relevant for the present analysis.
We assume the most general form of the superpotential, which is given
by  
\begin{align}
W &=\,
\wh{Q}\, 
(Y_1^d\wh{H}_1+Y_3^d\wh{H}_3+Y_5^d\wh{H}_5)\wh{D}^c+
\wh{L}\,(Y_1^e\wh{H}_1+Y_3^e\wh{H}_3+Y_5^e\wh{H}_5)\wh{E}^c
 \nonumber \\ 
& +\, \wh{Q}\,
(Y_2^u\wh{H}_2+Y_4^u\wh{H}_4+Y_6^u\wh{H}_6)\wh{U}^c+\mu_{12}\wh{H}_1\wh{H}_2+
\mu_{14}\wh{H}_1\wh{H}_4+\mu_{16}\wh{H}_1\wh{H}_6 \nonumber
\\ & + \,\mu_{32}\wh{H}_3\wh{H}_2+\mu_{34}\wh{H}_3\wh{H}_4+
\mu_{36}\wh{H}_3\wh{H}_6+\mu_{52}\wh{H}_5\wh{H}_2+\mu_{54}\wh{H}_5\wh{H}_4+
\mu_{56}\wh{H}_5\wh{H}_6\,,
\end{align}
where $\wh{Q}$ and $\wh{L}$ denote the quark and lepton $SU(2)_L$ 
doublet superfields, 
$\wh{U}^c$ and $\wh{D}^c$ are quark singlets, and $\wh{E}^c$ the lepton
singlet. The Yukawa matrices associated with each
Higgs superfield, $Y_i^q$, have been already defined in
Eq.~(\ref{Yd:1:5}). In what follows, we take
the $\mu_{ij}$ as effective parameters. 
(Notice that in this context the Giudice-Masiero 
mechanism~\cite{Giudice:1988yz} to generate the $\mu$-term through 
the K\"ahler potential is not available for prime orbifolds such as the $Z_3$
orbifold~\cite{LopesCardoso:1994is,Antoniadis:1994hg}.) 

The scalar potential receives the usual
contributions from $D$-, $F$- and SUSY soft-breaking terms, which we
write below, using for simplicity doublet components.
\begin{align}
V_F\,
=&
\operatornamewithlimits{\sum}_{\begin{smallmatrix}
{i,j=1,3,5}\\{l=2,4,6}
\end{smallmatrix}} \mu^*_{il}\, \mu_{jl}\, H_i^\dagger\, H_j
+
\operatornamewithlimits{\sum}_{\begin{smallmatrix}
{i=1,3,5}\\{k,l=2,4,6}
\end{smallmatrix}} \mu^*_{il} \,\mu_{ik}\, H_k^\dagger H_l \,,\nonumber\\
V_D\,
=&\,
\frac{g^2}{8} \,\operatornamewithlimits{\sum}_{a=1}^{3} 
\left[\, \operatornamewithlimits{\sum}_{i=1}^6
H_i^\dagger \,\tau^a\, H_i
\right]^2 + \,
\frac{g^{\prime 2}}{8} 
\left[\,\operatornamewithlimits{\sum}_{i=1}^6 \,
(-1)^i\, \left|H_i\right|^2 \,\right]^2\,,\nonumber\\
V_{\text{soft}}\,=&
\operatornamewithlimits{\sum}_{i,j=1,3,5}
(m^2_d)_{ij} \, H_i^\dagger\, H_j
+
\operatornamewithlimits{\sum}_{k,l=2,4,6}
(m^2_u)_{kl} \, H_k^\dagger H_l\,
-\operatornamewithlimits{\sum}_{\begin{smallmatrix}
{i=1,3,5}\\{j=2,4,6}
\end{smallmatrix}}
\left[(B\mu)_{ij}\, H_i\, H_j +\text{H.c.}\right]\,.\label{VDVFVS}
\end{align}

After electroweak symmetry breaking, the neutral components of the six
Higgs doublets develop VEVs, which we assume to be real, 
\begin{equation}
\vev{h^0_{1(3,5)}}\,=\,w_{1(3,5)}\,, 
\quad \quad \quad 
\vev{h^0_{2(4,6)}}\,=\,w_{2(4,6)}\,,
\end{equation}
and as usual one can write
\begin{align}
h^0_i \to w_i + \frac{1}{\sqrt{2}} \left(\sigma_i + i \varphi_i\right)\,.
\end{align}
For the purpose of minimising the Higgs potential and computing the
tree-level Higgs mass matrices, it proves more convenient to work in
the so-called ``Higgs basis''~\cite{Georgi:1978ri,Drees:1988fc}, 
where only two of the rotated fields develop VEVs:
\begin{align}\label{higgs:Ptransf}
&\quad \quad \quad \phi_i = P_{ij} h_j\,,\nonumber\\
& \vev{\phi^0_1} =\,  v_d\,, \quad \quad
\vev{\phi^0_2} =\,  v_u\,.
\end{align}
By construction (cf. Eq.~(\ref{ewz})), the new VEVs must satisfy 
\begin{equation}\label{ewz2}
v_u^2+v_d^2 \,\approx \, (174\,\text{ GeV})^2\,,
\end{equation}
and we can now define $\tan \beta$ (see Eq.(\ref{tb})) in the standard way,
\begin{equation}\label{tb2}
\tan \beta = \frac{v_u}{v_d}\,.
\end{equation}
In the new basis, the free parameters at the EW scale are 
$m^2_{ij}$, $b_{ij}$, which has dimensions mass$^2$, 
and $\tan \beta$ (for a detailed discussion of the Higgs basis, including
the definition of the new parameters and of $P_{ij}$, 
see~\cite{Escudero:2005hk}), and the minimisation
equations simply read:
\begin{equation}\label{minima:du}
\begin{array}{ll}
m^2_{11}\,=\,\,
b_{12} \,\tan \beta - \frac{M_Z^2}{2}\,\cos 2 \beta\,, \quad \quad
\quad \quad \quad
&
m^2_{22}\,=\,\,
b_{12} \,\cot \beta + \frac{M_Z^2}{2}\,\cos 2 \beta\,, \\
m^2_{13}\,=\,\,
b_{32} \,\tan \beta \,,
&
m^2_{24}\,=\,\,
b_{14}\, \cot \beta\,, \\
m^2_{15}\,=\,\,
b_{52} \, \tan \beta \,,
&
m^2_{26}\,=\,\,
b_{16} \,\cot \beta \,.
\end{array}
\end{equation}
After minimising the potential\footnote{We have verified that for each
  of the configurations analysed in this work, we are
  indeed in the presence of a local minimum with respect to the
  neutral Higgs scalars. Not only have we imposed the conditions for
  an extremum, but we also verified that it was a minimum by checking
  that all the minors of the Hessian matrices were positive
  definite. In terms of the Higgs spectrum, this is reflected in the
  absence of charged and neutral tachyonic states. Nevertheless, we do
  not discard the possibility of a global minimum associated with
  non-vanishing VEVs for the charged components of the six Higgs
  doublets.}, one can derive the charged, neutral
scalar and pseudoscalar mass matrices, and obtain the mass
eigenstates and the diagonalisation matrices. 
For the neutral states (scalars and pseudoscalars), the relation of mass and
interaction eigenstates is given by 
\begin{align}\label{higgsmass:diag}
S_R \,\mathcal{M}^2_R \,S_R^\dagger &= \Delta_R^2 = 
\operatorname{diag}({m^s_i}^2)\,, \quad i=1,\ldots,6\,, 
\nonumber\\
S_I \,\mathcal{M}^{2}_I \,S_I^\dagger &= \Delta_I^2=
\operatorname{diag}({m^p_i}^2)\,, \quad i=1,\ldots,6\,,
\end{align}
with $\Delta_{R,I}^2$ the diagonal scalar and pseudoscalar
squared mass eigenvalues (notice that the $i=1$ term for the
pseudoscalars corresponds to the unphysical massless would-be
Goldstone boson). We recall here that we are working in the
Higgs-basis, and that 
the matrices that diagonalise the mass matrices in the original basis
can be related to the latter as 
\begin{equation}\label{S:SRIP}
S_{\sigma,\varphi} \,=\,S_{R,I} \,P\,,
\end{equation} 
where
$P$ is the matrix appearing in Eq.~(\ref{higgs:Ptransf}). A detailed
study of such an extended Higgs sector, including choice of
basis, minimisation of the potential and derivation of the
tree-level mass matrices can be found in~\cite{Escudero:2005hk}.
It is also important to notice that throughout the analysis, and since
our aim is to investigate to which extent FCNCs push the lower
bounds on the Higgs masses, we do not consider radiative corrections
to the Higgs masses, using the bare masses instead. 

\section{Yukawa interactions and tree-level FCNCs}\label{yukint}
In the quark and Higgs mass eigenstate basis, the Yukawa interaction
Lagrangian reads:
\begin{align}
\mathcal{L}_{\text{Yukawa}}=&  
-\frac{1}{\sqrt{2}}
\operatornamewithlimits{\sum}_{i=1,3,5} \left[\,
\left(\mathcal{V}_d\right)^{ij}_{ab} \,
h_j^s \, \bar d_{R}^a \,d_{L}^b + 
i \left(\mathcal{W}_d\right)^{ij}_{ab} \,h_j^p \,\bar d_{R}^a \,d_{L}^b +
\text{H.c.} \right]\nonumber\\
&
-\frac{1}{\sqrt{2}}
\operatornamewithlimits{\sum}_{i=2,4,6} \left[
\left(\mathcal{V}_u\right)^{ij}_{ab} \,
h_j^s \, \bar u_{R}^a \,u_{L}^b + 
i \left(\mathcal{W}_u\right)^{ij}_{ab} \,h_j^p \,\bar u_{R}^a \,u_{L}^b +
\text{H.c.} \right]\,.
\end{align}
In the above, $a,b$ denote quark flavours, while
$i,j=1,\ldots,6$ are Higgs indices, with $s$ ($p$)
denoting scalar (pseudoscalar) mass eigenstates. The latter are related
to the original states as $h^s=S_\sigma \sigma$, $h^p=S_\varphi
\varphi$, as from Eqs.~(\ref{higgsmass:diag},\ref{S:SRIP}). 
The scalar (pseudoscalar) coupling
matrices $\mathcal{V}$ ($\mathcal{W}$) are defined as
\begin{align}\label{WV}
\left(\mathcal{V}_q\right)^{ij}_{ab} \,=\,&
(S_\sigma^\dagger)_{ij} \, \,
(V_R^q\,\, Y^{q}_i \,\,V^{q\dagger}_L)_{ab}\,, \nonumber \\
\left(\mathcal{W}_q\right)^{ij}_{ab} \,=\,&
(S_\varphi^\dagger)_{ij}\, \, 
(V_R^q \,\, Y^{q}_i \,\,V^{q\dagger}_L)_{ab}\,,
\end{align}
with $i=1,3,5 \ (2,4,6)$ for $q=d \,(u)$ and $j=1, \ldots, 6$.
$Y^q$ denote the Yukawa couplings, whose down-type elements were
displayed in Eq.~(\ref{Yd:1:5}), and $V_{L,R}$ are the
unitary matrices that diagonalise the quark mass matrices as 
\begin{equation}\label{quarkmass:diag}
V_R^q \,\, \mathcal{M}^q \, \,{V_L^q}^\dagger \,=\, 
\operatorname{diag}\,(m^q_i)\,, 
\quad \quad q=u,d\,,
\end{equation}
so that the Cabibbo-Kobayashi-Maskawa matrix is defined as
\begin{equation}\label{vckm:def}
V_{\text{CKM}}=V_L^u\, {V_L^d}^\dagger\,. 
\end{equation}
We emphasise that the matrices 
$V_{L,R}$ which diagonalise the quark mass matrices do not, in
general, diagonalise the corresponding Yukawa couplings. Hence, both
scalar and pseudoscalar Higgs-quark-quark interactions may exhibit a
strong non-diagonality in flavour space, which in turn translates in
the appearance of FCNCs and CP violation at the tree-level.

Even though a detailed discussion of FCNCs in multi-Higgs doublet
models was presented in~\cite{Escudero:2005hk}, we summarise here 
some relevant points, focusing on the neutral kaon sector and
investigating the tree-level contributions to $\Delta m_K$.
The latter is simply defined as the mass difference between the long-
and short-lived kaon masses,
\begin{equation}
\Delta m_K = m_{K_L} - m_{K_S} \simeq 2 \left|
\mathcal{M}^K_{12}\right|\,.
\end{equation} 
The contribution to $\mathcal{M}^K_{12}$ associated with the exchange
of scalar Higgses (with masses $m^s_j$) is given by 
\begin{align}\label{MK12:sigma}
\left. \mathcal{M}^K_{12}\right|^\sigma \,=\,&\,
\frac{1}{8} \operatornamewithlimits{\sum}_{\begin{smallmatrix}
{j=1-6}
\end{smallmatrix}} \frac{1}{(m^s_j)^2}
\left\{ \left[\operatornamewithlimits{\sum}_{i=1,3,5} \left(
{\mathcal{V}_d}^{ij*}_{12}+{\mathcal{V}_d}^{ij}_{21} \right)
\right]^2 \langle \overline K^0\left| (\bar s d) (\bar s d)
\right| K^0 \rangle \right. \nonumber\\
&+
\left.
\left[\operatornamewithlimits{\sum}_{i=1,3,5} \left(
{\mathcal{V}_d}^{ij*}_{12}-{\mathcal{V}_d}^{ij}_{21} \right)
\right]^2 \langle \overline K^0\left| (\bar s \gamma_5 d) (\bar s
\gamma_5 d)
\right| K^0 \rangle
\right\}\,, 
\end{align}
while the exchange of a pseudoscalar state (with mass $m^p_j$) reads
\begin{align}\label{MK12:phi}
\left. \mathcal{M}^K_{12}\right|^\varphi \,=\,&\,
\frac{1}{8} \operatornamewithlimits{\sum}_{\begin{smallmatrix}
{j=2-6}
\end{smallmatrix}} \frac{1}{(m^p_j)^2}
\left\{ \left[\operatornamewithlimits{\sum}_{i=1,3,5} \left(
{\mathcal{W}_d}^{ij*}_{21}-{\mathcal{W}_d}^{ij}_{12} \right)
\right]^2 \langle \overline K^0\left| (\bar s d) (\bar s d)
\right| K^0 \rangle \right. \nonumber\\
&+
\left.
\left[\operatornamewithlimits{\sum}_{i=1,3,5} \left(
{\mathcal{W}_d}^{ij*}_{21}+{\mathcal{W}_d}^{ij}_{12} \right)
\right]^2 \langle \overline K^0\left| (\bar s \gamma_5 d) (\bar s
\gamma_5 d)
\right| K^0 \rangle
\right\}\,. 
\end{align}
Once all the contributions to $\mathcal{M}^K_{12}$ have been taken
into account, the prediction of this orbifold model
regarding $\Delta m_K$ should be compared with the experimental value, 
$(\Delta m_K)_{\text{exp}} \simeq 3.49 \times 10^{-12}$ MeV~\cite{pdg2004}.
For the other neutral meson systems, $B_d$, $B_s$ and $D^0$, the computation is
analogous. In each case the viability of the model imposes that the
obtained results should be compatible with the current bounds:
$(\Delta m_{B_d})_{\text{exp}} \simeq 3.304 \times 10^{-13}$ GeV, 
$(\Delta m_{B_s})_{\text{exp}}>94.8 \times 10^{-13}$ GeV and 
$\Delta m_{D^0} < 46.07 \times 10^{-12}$ MeV~\cite{pdg2004}.

Before proceeding to the numerical analysis, let us briefly comment on
the several contributions.
First of all, it is widely recognised that, in models with tree-level
FCNCs, the most stringent bounds are usually associated with $\Delta
m_K$. For the $Z_3$ orbifold scenario, with hierarchical Yukawa
couplings, one expects the bound from $\Delta m_{B_d}$ to be less
severe than that of $\Delta m_K$. The same should occur for the $B_s$
mass difference, since in the SM this mixing is already maximal (the only
exception occurring if new contributions matched exactly those of
the SM, but had opposite sign, in which case a cancellation could take
place).
The $D^0$ mass difference can be quite challenging to accommodate.
As pointed out in~\cite{McWilliams:1980kj,Cheng:1987rs} 
and~\cite{D:burdman:datta}, models allowing for FCNC at the tree-level
may present the possibility of very
large contributions to $\Delta m_D$, and the latter could even exceed 
by a factor 20 those to $\Delta m_K$~\cite{Cheng:1987rs}. 
Nevertheless, one should bear in mind the
fact that mixing in the $D^0$ sector is very sensitive
to the hadronic model used to estimate the transition amplitudes, and
there is still a very large uncertainty in deriving its decay
constants, etc. Therefore, the constraints on a given model arising
from $\Delta m_D$ should not be over-emphasised, and we will adopt
this conservative view throughout our discussion.

Another interesting issue\footnote{Extended Higgs sectors with flavour
  violation have other interesting consequences, such as flavour
  violating Higgs and top-decays. For a discussion of the latter, and
  the associated experimental signatures at the next generation of
  colliders, see, for
  example~\cite{Bejar:2005kv,Curiel:2003uk,Aguilar-Saavedra:2004wm},
  and references therein.} is that of rare decays. It has been argued  
that, again when no theory for the full Yukawas is available,
some rare decays become very sensitive to flavour changing 
contributions induced by Higgs exchange at the
tree-level~\cite{Sher:1991km}.
In the present model, the Yukawas are well-defined, not only for the
quark, but also for the lepton sector. In a forthcoming
work~\cite{EJMT}, we will
analyse in detail the lepton sector of this class of orbifold
constructions, taking also into account the potentially most
constraining decay modes, as $\mu \to e \gamma$, $B_d \to K \mu \tau$
and $B_s \to \mu \tau$. 

Additionally, and given the existence of flavour violating 
neutral Higgs couplings, and
the possibility of having complex Yukawa couplings, 
it is natural to have tree-level contributions to CP violation. In the
kaon sector, indirect CP violation is parameterised by
$\varepsilon_K$, and defined as 
\begin{equation}
\varepsilon_K \,=\, -\frac{e^{i \pi/4}}{\sqrt{2}} \, 
\frac{\operatorname{Im} \left[\mathcal{M}^K_{12} \,
\lambda_u^2\right]}{|\lambda_u|^2 \, \Delta m_K}\,,
\end{equation}
where $\lambda_u$ is defined from CKM elements as $\lambda_u= V^*_{us}
V_{ud}$. From experiment one has 
$\varepsilon_K = (2.284 \pm 0.014) \times 10^{-3}$~\cite{pdg2004}.
In this case, and since we are in the presence of tree-level, rather
than 1-loop interactions, the new contributions to $\varepsilon_K$
are expected to be quite large, even if the amount of CP 
violation associated with the CKM matrix (for instance parameterised by
$J_{\text{CP}}$~\cite{Jarlskog:1985ht}) 
is far smaller than the value derived
from usual SM fits - $J_{\text{CP}} \sim \mathcal{O}(10^{-5})$~\cite{pdg2004}.

\section{Numerical results}\label{results}
In the present scenario, most of the observables addressed in the previous 
Section receive their dominant contributions from tree-level
processes. This situation strongly diverges from the usual scenarios
of both SM and MSSM, where FCNCs only occur at the 1-loop level.
Given the increasing experimental accuracy, it is important to
investigate to which extent the present scenario is compatible with
current experimental data.

We divide the numerical approach in two steps.
Firstly, we focus on the string sector of the model, and for each
point in the space generated by the free parameters of the orbifold
($\varepsilon_5, \alpha^f$), 
we derive the up- and down-quark mass matrices\footnote{
It is worth emphasising here that the quark masses appearing 
in Eq.~(\ref{vev:quarkmass}) (and in all subsequent relations) are used
on the sole purpose of obtaining an approximate determination of the
VEVs. Afterwards, the actual values of $m_{q_i}$ are exactly computed.}
and compute the CKM matrix.
This procedure allows us to investigate the several regimes of
parameters that translate into viable quark spectra, and discuss the
implications of the relations between the several parameters. At this
early stage, we consider only real values for the orbifold parameters.
Further imposing the conditions associated with EWSB 
given in Eq.~(\ref{wmz}), and fixing a 
value\footnote{We recall here that $\tan \beta$ is a necessary
  parameter to specify the Higgs sector, which in turn is mandatory to
  investigate the issue of FCNCs in the present model.} for $\tan
\beta$, one can
then determine the values of $g\,N$ and 
$\varepsilon_1$ (cf. Eqs.~(\ref{wmbeta},\ref{auad:rel})).
Another possible approach would be to scan over the space generated by
the moduli ($T_i$) and the VEVs of the $SU(3) \times SU(2)$ singlet
fields ($c_i$), but this would translate in less
straightforward relations between the orbifold parameters and the experimental 
data.
A secondary step requires specifying the several Higgs parameters,
which must obey the minimum criteria of Eq.~(\ref{minima:du}). 
Finally, the last step comprehends the 
analysis of how each of the Yukawa patterns constrains
the Higgs parameters in order to have compatibility with the FCNC
data. In particular, we want to investigate how heavy the scalar and
pseudoscalar eigenstates are required to be in order 
to accommodate the observed
values of $\Delta m_K$, $\Delta m_{B_d}$, etc.

\subsection{Quark Yukawa couplings and the CKM matrix}\label{quark:orbifold}
As discussed in~\cite{Abel:2002ih}, there are three regimes for the
values of $\alpha^f $ and $\beta^f $, depending on the specific
orbifold configuration :

\noindent \hspace*{40mm}
(a)\, $\alpha^f \sim \varepsilon_5$ and $\beta^f  \sim 1$; 

\noindent\hspace*{40mm}
(b)\, $\alpha^f  \approx \varepsilon_5$ and 
$\beta^f  \approx 1/\varepsilon_5$; 

\noindent\hspace*{40mm}
(c)\, $\alpha^f  \sim 1$ and $\beta^f  \sim 1$.

\noindent
In any case, it is clear from Eq.~(\ref{alpha:beta:def}) that 
$\alpha^f$ and $\beta^f$ must obey, by construction, the
following bounds: 
\begin{equation}
\varepsilon_5 \lesssim \alpha^f \lesssim 1\,,
\quad \quad \quad 
1 \lesssim \beta^f \lesssim \frac{1}{\varepsilon_5}\,.
\end{equation} 
In what follows we investigate whether each point in the orbifold
parameter space can be associated with a consistent quark spectrum and mixings.
For given values of the input quark masses, one fixes the ratio of the
several Higgs VEVs, which in turn allows to reconstruct the full quark
mass matrices, and obtain the mass eigenstates and CKM matrix.
In particular, throughout this analysis we shall focus on four sets of
input quark masses, whose values are listed in Table~\ref{set:ae}.
\begin{center}
\begin{table}\hspace*{35mm}
\begin{tabular}{|c|cccccc|}
\hline
Set & $m_u$ & $m_d$ & $m_c$ & $m_s$ & $m_t$ & $m_b$ \\
\hline
A & 0.004  & 0.008 & 1.35 & 0.13  & 180 & 4.4 \\
B & 0.0035 & 0.008 & 1.25 & 0.1   & 178 & 4.5 \\
C & 0.0035 & 0.004 & 1.15 & 0.08  & 176 & 4.1 \\
D & 0.004  & 0.006 & 1.2  & 0.105 & 178 & 4.25 \\
\hline
\end{tabular}
\caption{Sets of input quark masses (in GeV) used in the numerical analysis.}
\label{set:ae}
\end{table}
\end{center}
Throughout, we require the CKM matrix elements to lie within the
following ranges~\cite{pdg2004}:
\begin{equation}\label{vckm:exp:limits}
V_\text{CKM}\,=\,\left(
\begin{array}{ccc}
0.9739 - 0.9751 & 0.221 - 0.227 & 0.0029 - 0.0045\\
0.221 - 0.227 & 0.9730 - 0.9744 & 0.039 - 0.044\\
0.0048 - 0.014 & 0.037 - 0.043 & 0.9990 - 0.9992
\end{array}\right)\,.
\end{equation}
Regarding the effect of the orbifold parameters on quark mixing,
let us notice that both $\varepsilon_5$ and $\alpha^{u^c}$ are
crucial in obtaining the Cabibbo angle. 
As expected, the down-sector parameters are those directly responsible
for the mixings between generations, and their role is particularly
relevant in determining $V_{td}$ ($\alpha^{d^c}$ - and to a lesser
extent, also $\alpha^{u^c}$), $V_{ts}$, 
$V_{ub}$ and  $V_{cb}$. Once these elements are fixed in
accordance with experiment, the others are usually also in agreement. 
Finally, let us recall that from choosing a concrete value for $\tan
\beta$, and complying with the bound on $M_Z$ from EW symmetry
breaking, Eqs.~(\ref{wmz},\ref{wmbeta}), one is implicitly fixing for
each set of $\{\varepsilon_5, \alpha^f, \beta^f\}$, the values of 
$\varepsilon_1$ and $g\,N$. 

\begin{figure}[t]
  \begin{center} \hspace*{-10mm}
    \begin{tabular}{cc}
	\psfig{file=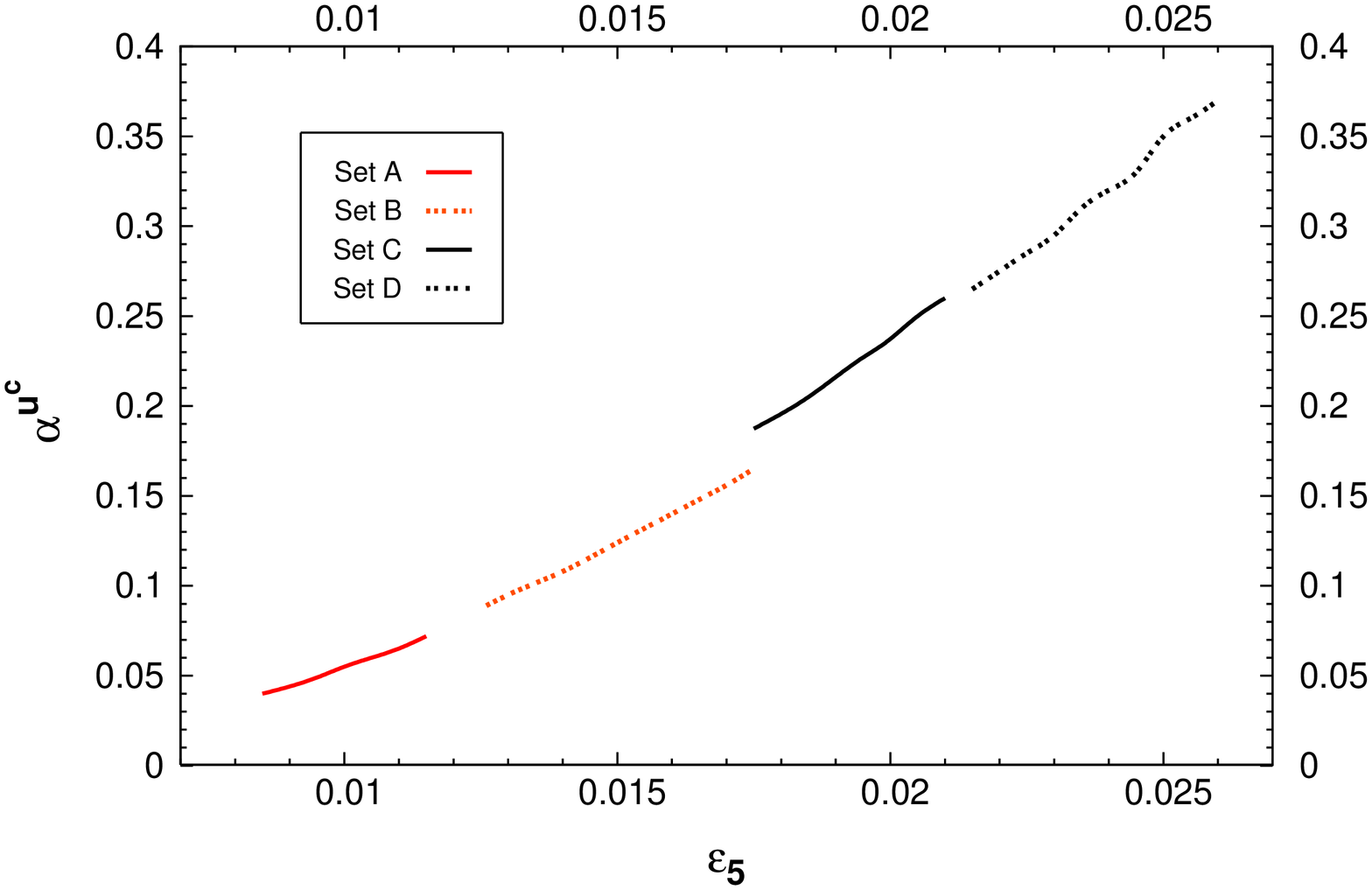,width=55mm,angle=270,clip=} &
	\psfig{file=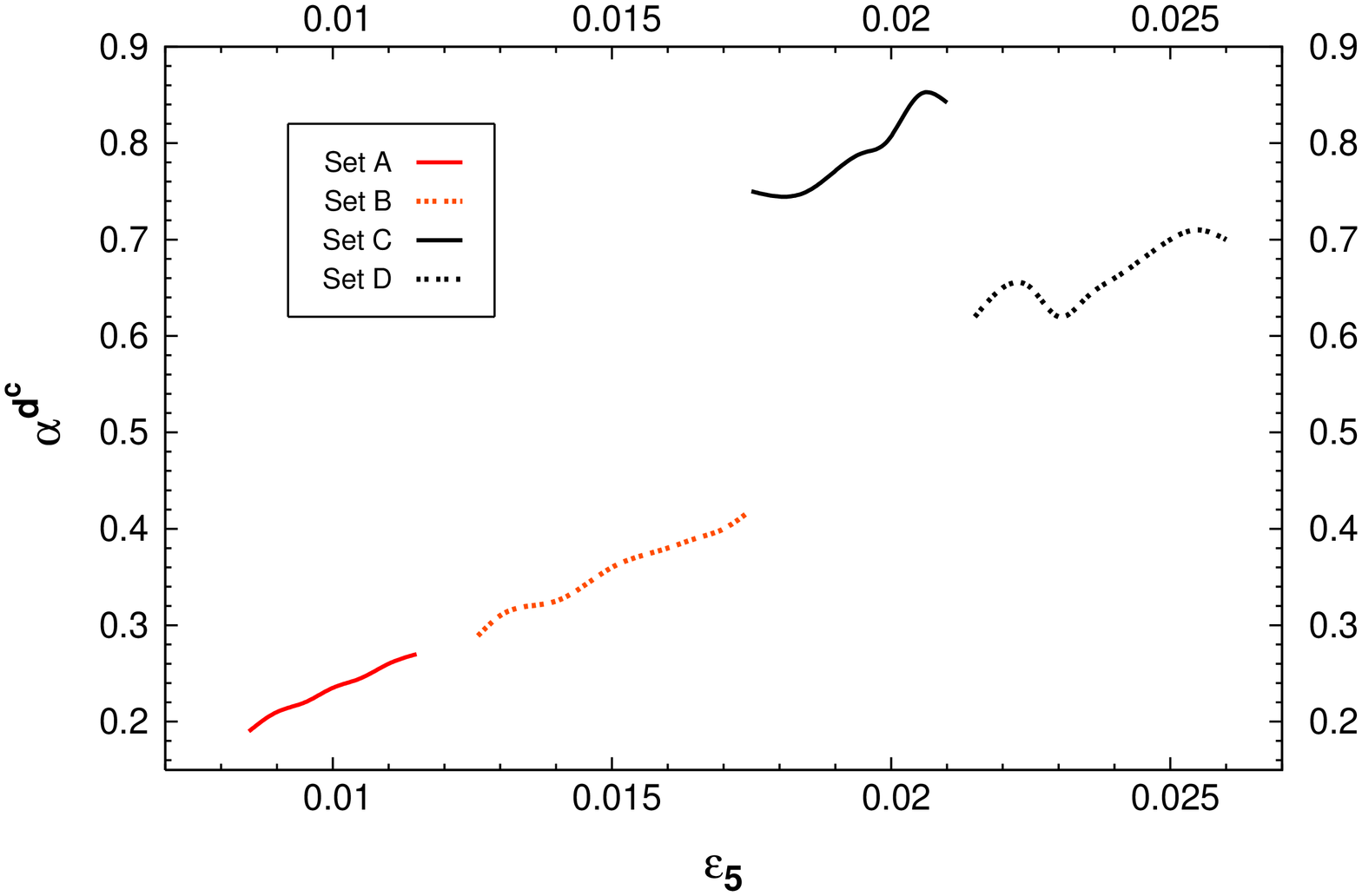,width=55mm,angle=270,clip=} 
    \end{tabular}
    \caption{Correlation between the orbifold parameters
($\alpha^{u^c}$, $\varepsilon_5$) and ($\alpha^{d^c}$,
    $\varepsilon_5$), for distinct sets of input quark masses, A-D
      (red full lines, red dashed lines, full black and dashed black
      lines, respectively).} 
    \label{fig:orbifold:alpha:ud:e5}
  \end{center}
\end{figure}

In Fig.~\ref{fig:orbifold:alpha:ud:e5}, we present the correlation between the
orbifold parameters, for the four sets of input quark masses given in 
Table~\ref{set:ae}.
We only present points that are associated with viable quark masses and that  
are in agreement with current bounds on $|V_{\text{CKM}}|$ (from
Eq.~(\ref{vckm:exp:limits})). 
Rather than scatter plots, in Fig.~\ref{fig:orbifold:alpha:ud:e5} we present
sets of points. This is due to having very narrow intervals of
fluctuation for all the parameters. As an example, let us mention that
for constant values of $\varepsilon_5$, $\alpha^{d^c}$ and $\beta^f$, 
$\alpha^{u^c}$ is fixed within a $2\%$ interval. 
>From Fig.~\ref{fig:orbifold:alpha:ud:e5} it is clear that for a given
set of Higgs VEVs (determined by the associated set of input quark
masses - Table~\ref{set:ae}) the allowed orbifold parameter space
is very constrained. This is a direct consequence of the increasing
accuracy in the experimental determination of the
$V_\text{CKM}$. 
In each set (A--D), moving outside the displayed ranges
would translate in violating the experimental bounds on (at least) one
of the CKM matrix elements. Larger values of $\varepsilon_5$ would
also (typically) be associated with an up-quark mass below the current
accepted range. 
Regarding the values of $\beta^f$, these are constrained to be 
$\beta^f \approx 1$ throughout the allowed parameter space
(cf. Eq.~(\ref{bf:alpha})).
 From the direct inspection of Fig.~\ref{fig:orbifold:alpha:ud:e5},
together with the fact that $\beta^f \approx 1$, 
it appears that at least two regimes for $\alpha^f$
are present. For the up sector, one would suggest that the orbifold
configuration is such that $\alpha^{u^c}$ is roughly 
$\mathcal{O}(\varepsilon_5)$, so that one is faced with regime (a).
The same would happen for sets A and B in the down sector, although 
sets C and D appear to favour a regime with $\alpha^{d^c} \sim 1$,
thus suggesting regime (c).

Let us now aim at understanding the behaviour of both $\alpha^{u^c,d^c}$ as a
function of $\varepsilon_5$. In Ref.~\cite{Abel:2002ih}, 
several analytical relations for the CKM matrix
elements as a function of $\{\varepsilon_5,\alpha^f,\beta^f\}$ were
derived. Although those relations were computed for the case of
hermitic mass matrices, where
$V^f_L = V^f_R$, and are thus not truly valid for the present case,  
they are quite useful in understanding
Fig.~\ref{fig:orbifold:alpha:ud:e5}. For instance, 
the Cabibbo angle is given, to a very good approximation, by
\begin{equation}\label{Vus:approx}
V_{us} \approx
-\varepsilon_5^2 \left(\frac{m_t}{m_c} \,\frac{1}{\alpha^{u^c}} -
\frac{m_b}{m_s} \,\frac{1}{\alpha^{d^c}}\right)\,,
\end{equation}
and the above expression 
is clearly dominated by the first term on the right-hand side
(r.h.s). From Eq.~(\ref{Vus:approx}), 
it becomes transparent that the dependence of 
$\alpha^{u^c}$ on $\varepsilon_5$ should indeed be parabolic, as
clearly displayed in Fig.~\ref{fig:orbifold:alpha:ud:e5}.
Regarding $\alpha^{d^c}$, its evolution is strongly influenced by the
allowed regions of $\varepsilon_5$ (as dictated by the quark mass
regimes taken as input, especially $m_s/m_b$). 
For sets A and B, the ratio of down-type quark masses is such that the
leading term in the analytical expression of $V_{ub}$,
\begin{align}\label{Vub:approx}
V_{ub} &\approx
\left(\frac{m_s}{m_b} \alpha^{d^c}-\frac{m_c}{m_t} \alpha^{u^c}
\right) 
- \varepsilon_5 \left(
\frac{m_d}{m_b} \frac{m_t}{m_c} \frac{\alpha^{d^c}}{\alpha^{u^c}
\beta^{d^c}} - \frac{m_u}{m_t} \frac{1}{\beta^{u^c}}\right) + 
\mathcal{O}(\varepsilon_5^6)
\,,
\end{align}
provides a reasonable understanding of $\alpha^{d^c}
(\varepsilon_5)$. For sets C and D, the situation is more involved,
and the second term on the r.h.s. of Eq.~(\ref{Vub:approx}) plays an
important role. In fact, $\alpha^{d^c}$ receives contributions which
display a near-resonant behaviour for the input quark mass ratios in
the correspondent $\varepsilon_5$ range. This is the origin of the
unexpected positioning of set C in Fig.~\ref{fig:orbifold:alpha:ud:e5}.

\begin{figure}[t]
  \begin{center} 
\psfig{file=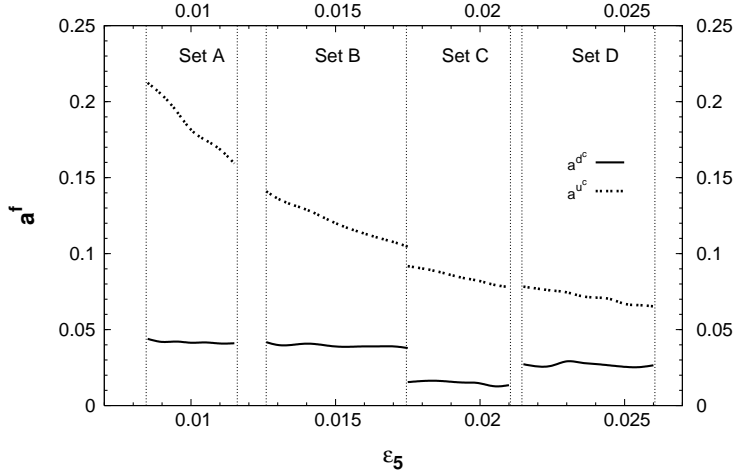,width=70mm,angle=270,clip=} 
    \caption{Values of $a^{d^c}$ and $a^{u^c}$ (full and dashed lines,
      respectively) as a function of $\varepsilon_5$. 
        Vertical dotted lines isolate the ranges of $\varepsilon_5$
	associated with each set A-D.} 
    \label{fig:orbifold:au:auTB:e5}
  \end{center}
\end{figure}

We now address the behaviour of $a^{u^c}$ and $a^{d^c}$.
By construction, and as clearly seen from Eq.~(\ref{af:alpha:beta}), once 
$\varepsilon_5$ and $\alpha^f$ (or equivalently $\varepsilon_5$ and
$\beta^f$) are set, one is implicitly fixing $a^{f}$. Moreover,
satisfying the EWSB conditions for the VEVs, together with imposing a
given value of $\tan \beta$ also translates in determining 
$g\,N$ and $\varepsilon_1$.
In Fig.~\ref{fig:orbifold:au:auTB:e5}, we plot the values of 
$a^{u^c}$ and $a^{d^c}$ as a function of $\varepsilon_5$, 
as determined from Eq.~(\ref{af:alpha:beta}).  
As seen from Fig.~\ref{fig:orbifold:au:auTB:e5}, 
the behaviour of set C regarding $a^{d^c}$ 
is slightly misaligned with what one would
expect from the analysis of sets A, B and D. Notice however that this is
due to the dependence of $a^{d^c}$ on $\alpha^{d^c}$ 
(cf. Eq.~(\ref{af:alpha:beta})). 
Recall from Fig.~\ref{fig:orbifold:alpha:ud:e5} that
for set C, the allowed values of $\alpha^{d^c}$ were somewhat higher 
than for the other cases, and this is the source of the suppression
displayed in Fig.~\ref{fig:orbifold:au:auTB:e5}, set C.

\begin{figure}[t]
  \begin{center} 
	\psfig{file=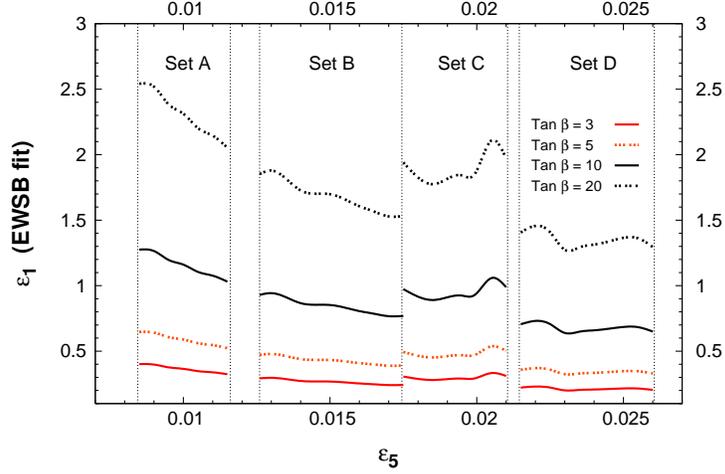,width=70mm,angle=270,clip=} 
    \caption{$\varepsilon_1$ as a function of $\varepsilon_5$
	for $\tan \beta=$3, 5, 10 and 20 (red full lines, red dashed
        lines, full black and dashed black lines, respectively).
	Vertical dotted lines isolate the ranges of $\varepsilon_5$
	associated with each set A-D.}
            \label{fig:orbifold:e1:adTB:e5}
  \end{center}
\end{figure}

In Fig.~\ref{fig:orbifold:e1:adTB:e5}, we present the values of 
$\varepsilon_1$ as a function of $\varepsilon_5$ for the four sets of
quark masses, and distinct regimes of $\tan \beta$.
>From Fig.~\ref{fig:orbifold:e1:adTB:e5} we can also verify that the values of
$\varepsilon_1$ are, in general, larger than those of $\varepsilon_5$. 
The ``misaligned'' behaviour of set C is again a consequence 
of the effects already discussed. 
Regarding the actual value of $g\,N$ it suffices to mention that for
the orbifold parameter space here investigated, and for the values of
$\tan \beta$ here selected, one has 
$1.03\,\lesssim g\,N \lesssim \,1.16$\,.

\begin{figure}[t]
  \begin{center} \hspace*{-10mm}
    \begin{tabular}{cc}
	\psfig{file=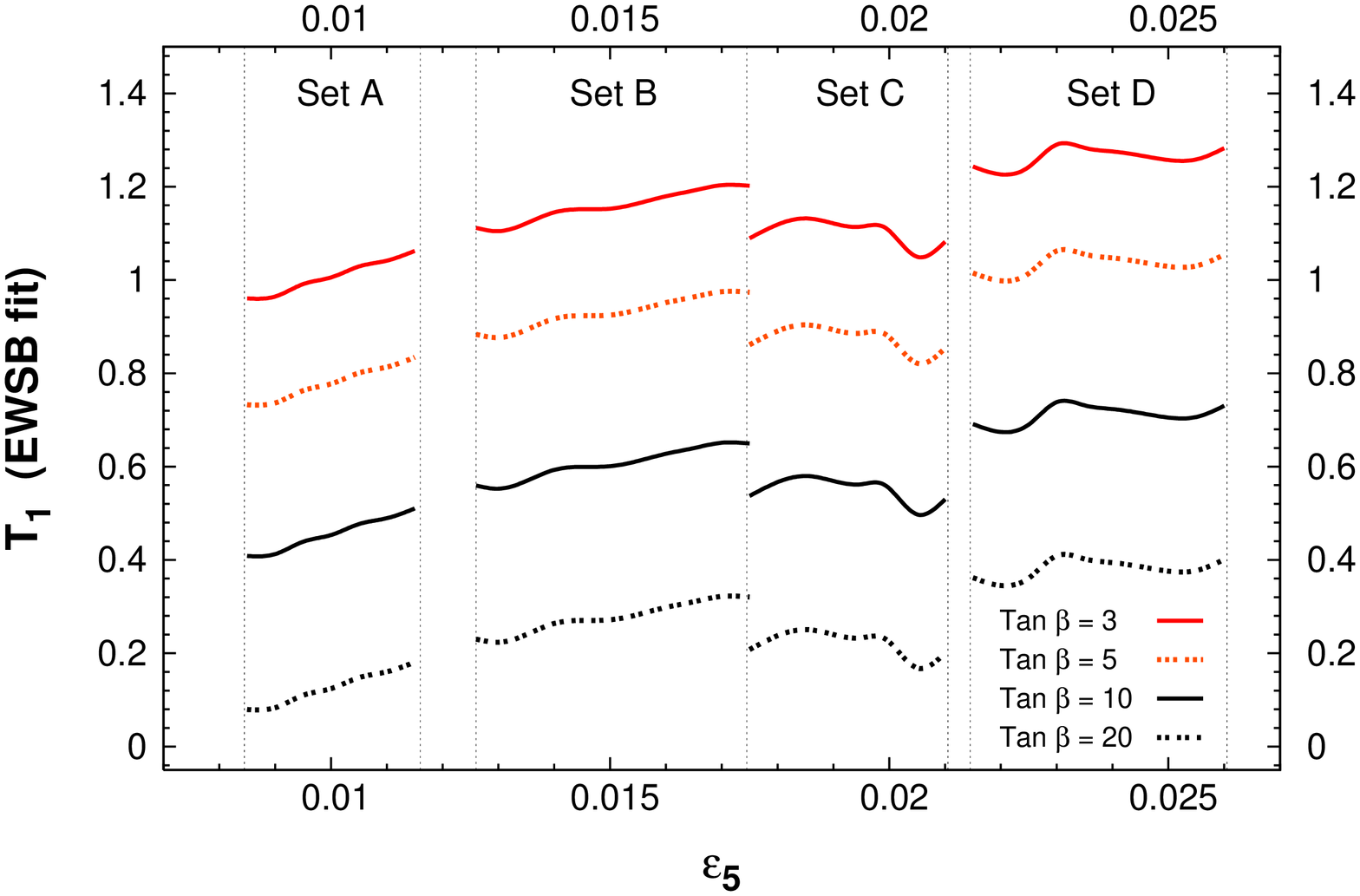,width=55mm,angle=270,clip=} &
	\psfig{file=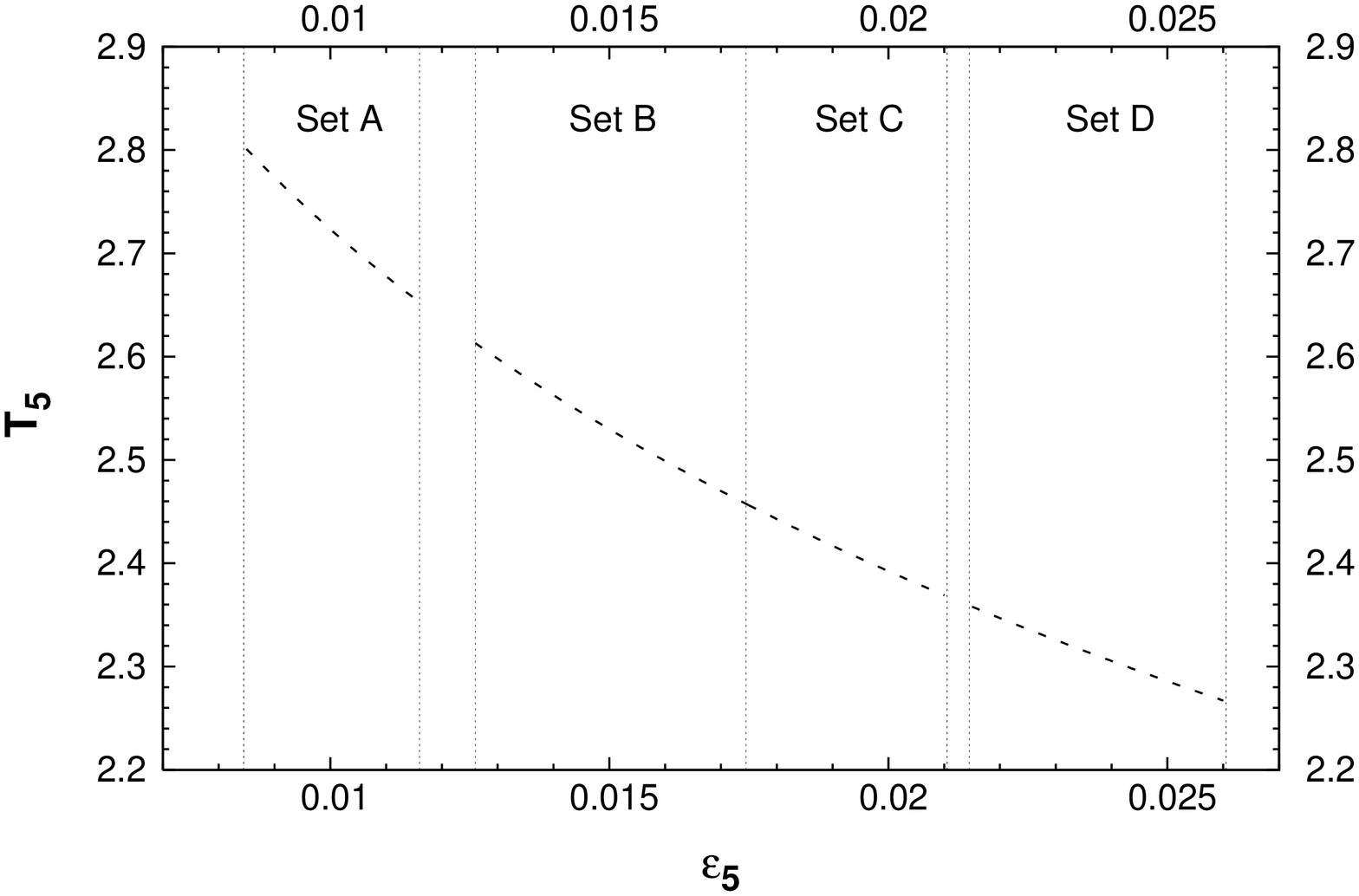,width=55mm,angle=270,clip=} 
    \end{tabular}
    \caption{Diagonal lattice moduli, $T_1$ (left) and $T_5$ (right), 
    as a function of $\varepsilon_5$. For the case of
    $T_1$, we plot it for 
    several values of $\tan \beta=$3, 5, 10 and 20 (following the line and
    colour scheme of Fig.~\ref{fig:orbifold:e1:adTB:e5}). For $T_5$ the
    dotted line denotes the prediction of the orbifold.} 
    \label{fig:orbifold:T5:T1:e5}
  \end{center}
\end{figure}

Since we now have the relevant information, we can further compute the
value of the lattice's diagonal moduli, $T_1$ and $T_5$, as defined in
Eq.~(\ref{eiTi}). The value of $T_5$ is unambiguously determined.
Nevertheless, and since the determination of $\varepsilon_1$ is a direct
consequence of complying with the EWSB conditions for a fixed value of 
$\tan \beta$, its determination varies accordingly.
In Fig.~\ref{fig:orbifold:T5:T1:e5} we display the diagonal moduli
as a function of $\varepsilon_5$, for $\tan \beta=$3, 5, 10 and 20. 
>From Fig.~\ref{fig:orbifold:T5:T1:e5}, it is manifest that for the
parameter space investigated, the values
of the diagonal moduli, $T_1$ and $T_5$, are never degenerate. This
confirms our original assumption (see footnote~\ref{Tifoot}) 
that distinct moduli
are indeed required to accommodate the experimental data. 
Although we have allowed for non-degenerate $T_i$, this remains quite
a restrictive choice. We recall that we still have six additional
degrees of freedom, namely the angles between the complex planes,
which we have taken as zero in the present analysis.

\begin{figure}[t]
  \begin{center} \hspace*{-10mm}
    \begin{tabular}{cc}
      \psfig{file=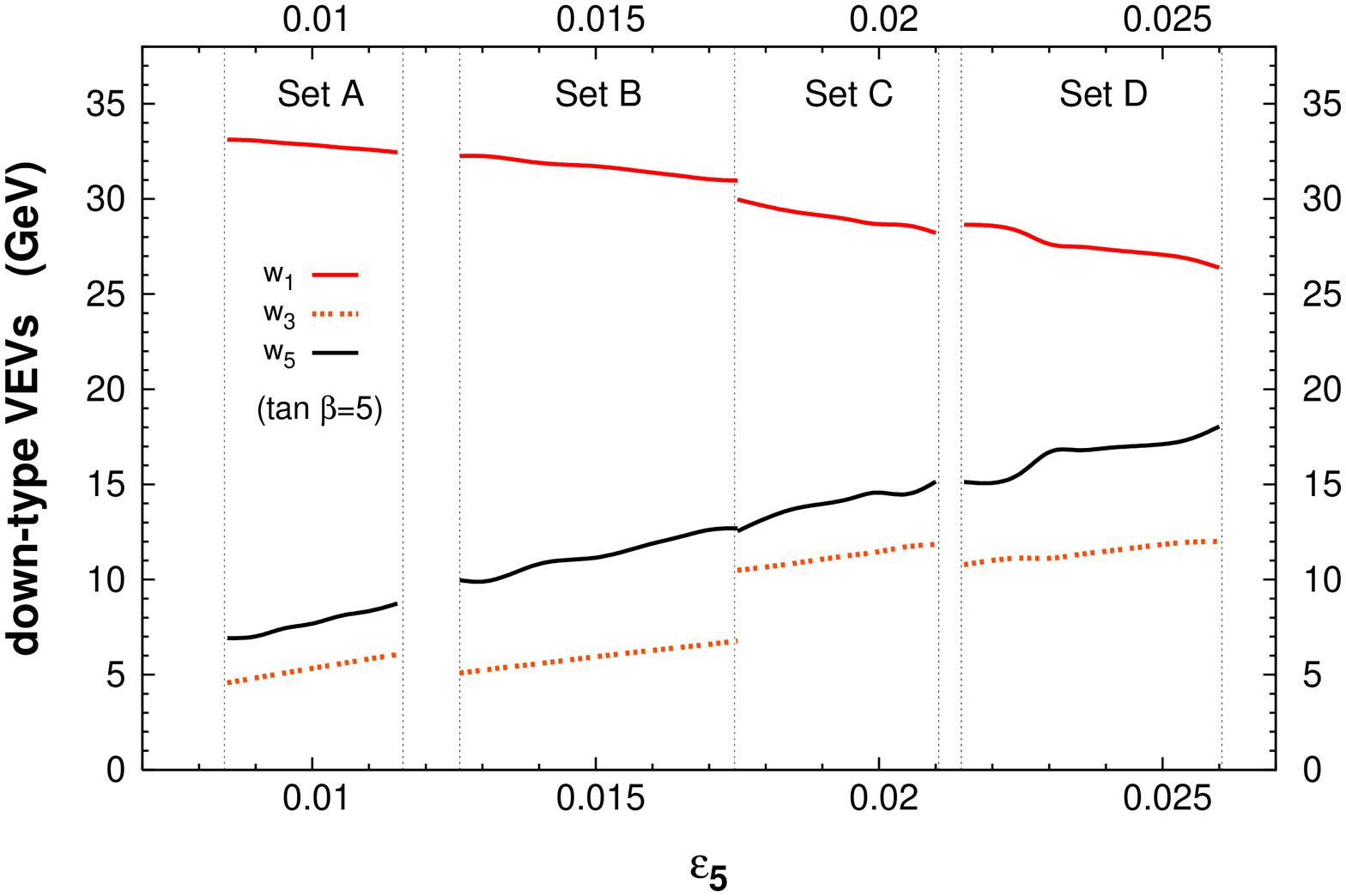,width=55mm,angle=270,clip=} &
      \psfig{file=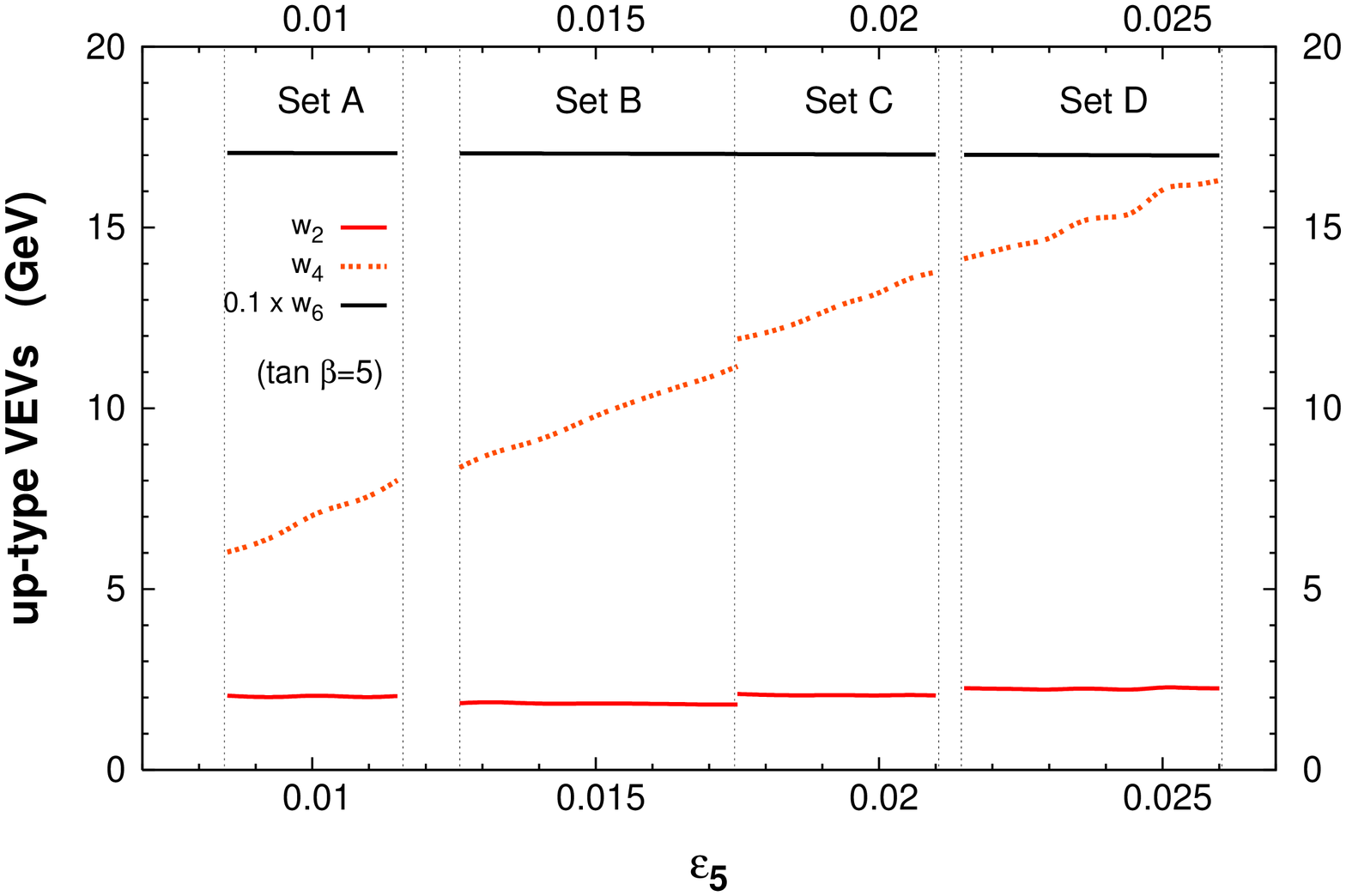,width=55mm,angle=270,clip=}
    \end{tabular}
    \caption{Down- and up-type Higgs VEVs, $w_{1,3,5}$ and
    $w_{2,4,6}$, as a function of $\varepsilon_5$, for $\tan
    \beta=$5. For convenience, we plot $w_6/10$. As before, vertical
    dotted lines isolate the ranges of $\varepsilon_5$ associated with 
    each set A-D.}
    \label{fig:orbifold:ctw:e5}
  \end{center}
\end{figure}

To conclude the study of the orbifold parameter space,
let us just plot the values of the Higgs
VEVs, again as a function of $\varepsilon_5$. As an illustrative
example, depicted in Fig.~\ref{fig:orbifold:ctw:e5}, we take $\tan \beta=5$.
It is interesting to comment that in the up-sector, the VEVs exhibit a
clearly hierarchical pattern, $w_2<w_4<w_6$ while in the down-type
VEVs we encounter a not so definite pattern, with 
$w_3<w_5<w_1$. This is a direct consequence of the relations 
given in Eq.~(\ref{vev:quarkmass}).

Finally, let us summarise our analysis of the orbifold parameter space
by commenting on the relative number of input parameters and number of
observables fitted. 
Working with the six Higgs VEVs ($w_i$), and the orbifold parameters 
$\varepsilon_1$, $\varepsilon_5$, $\alpha^{u^c}$ and $\alpha^{d^c}$,
one can obtain the correct EWSB ($M_Z$), as well as the correct quark
masses and mixings (six masses and three mixing angles).

\subsection{Tree-level FCNCs in neutral mesons}
The present orbifold model does not include a specific prediction 
regarding the Higgs sector. For instance, we have no
hint regarding the value of the several bilinear terms, nor
towards their origin. 
Concerning the soft breaking terms, the situation is identical. Since
the FI $D$-term, which could have broken SUSY at the string scale, was
cancelled, one must call upon some other mechanism to ensure that
supersymmetry is indeed broken in the low-energy theory.
In the absence of further information, we merely assume that the 
structure of the soft
breaking terms is as in Eq.~(\ref{VDVFVS}), taking the Higgs soft
breaking masses and the $B \mu$-terms as free parameters (provided
that the EW symmetry breaking and minimisation conditions are verified).

Before proceeding, it is important 
to stress that the Higgs spectrum (i.e. the scalar and
pseudoscalar physical masses) cannot be a direct input when
investigating the occurrence of FCNCs. In some previous studies of FCNCs in
multi-Higgs doublets models (see for example~\cite{Cheng:1987rs}) 
the bounds
were derived for the diagonal entries in the scalar and pseudoscalar
mass matrices. However, this
approach neglects mixings between the several fields $R_i, I_i$, 
and excludes the exchange of some scalar and pseudoscalar states.

Although it is possible
to begin the analysis from the original basis (where all neutral Higgses
develop VEVs), 
we prefer to define the Higgs parameters on the Higgs-basis, relying on
the minima conditions (and associated inequalities) to ensure that we
are on the presence of true minima.
Therefore, the parameters that must be specified are:
\begin{equation}
\tan \beta\,, \quad m^2_{ij}\,, \quad b_{ij}\,,
\end{equation}
entering in Eq.~(\ref{minima:du}).
In the absence of orbifold predictions for the Higgs sector
parameters, and 
motivated by an argument of simplicity, we begin our analysis by
considering textures for the above parameters as simple
as possible. 

To avoid rewriting the Higgs soft-breaking
masses, we adopt the following parameterisation: 
\begin{equation}\label{higgs:tx:mud}
m^{(d)}_{ij} = \left(
\begin{array}{ccc}
\otimes&\otimes  &\otimes  \\
\otimes & x_3 & y \\
\otimes & y & x_{5}
\end{array}\right)\times {1 \text{TeV}}\,,
\quad 
m^{(u)}_{kl}= \left(
\begin{array}{ccc}
\otimes&\otimes  &\otimes  \\
\otimes & x_4 & y \\
\otimes & y & x_{6}
\end{array}\right)\times {1 \text{TeV}}\,,
\quad
\sqrt{b_{ij}}=b \times {1 \text{TeV}}\,.
\end{equation}
In the above, $m^{d(u)}$ should be understood as the $i,j=1,3,5$
($k,l=2,4,6$) submatrices of the $6\times6$ matrix that encodes the rotated
soft-breaking Higgs masses in the Higgs basis (see~\cite{Escudero:2005hk}).
The symbol $\otimes$ denotes an entry which is
fixed by the minima equations of Eq.~(\ref{minima:du}). 
This parametrisation allows to easily define the Higgs sector
via six dimensionless parameters.
We begin by taking a near-universality limit for the Higgs-sector
textures introduced in Eq.~(\ref{higgs:tx:mud}). 
Regarding the value of $\tan \beta$,
and if not otherwise stated, we shall take $\tan \beta=5$ in the subsequent
analysis. 
We first consider the following four cases, presenting the associated
scalar and pseudoscalar Higgs spectra:
\begin{itemize}
\item[(a)] 
$x_3=x_4=0.5\,, x_5=x_6=0.75\,, y=0.1\,, b=0.1\,$

$m^s = \{82.5, 190.6, 493.9, 515.9, 744.4, 760.2\}$ GeV\,;

$m^p = \{186.8, 493.9, 515.9, 744.4, 760.2\}$ GeV\,.
\item[(b)] 
$x_3=x_4=0.4\,, x_5=x_6=0.8, y=0.15\,, b=0.2\,$

$m^s = \{84.6, 213.9, 387.4, 560.8, 785.9, 879.1\}$ GeV\,;

$m^p = \{215.2, 387.3, 560.5, 785.9, 878.9\}$ GeV\,.
\item[(c)] 
$x_3=x_4=0.75\,, x_5=x_6=1, y=0.25\,, b=0.2\,$

$m^s = \{83.6, 292.9, 733.6, 785.9, 987.6, 1057.0\}$ GeV\,;

$m^p = \{291.1, 733.6, 785.9, 987.6, 1057.0\}$ GeV\,.
\item[(d)] 
$x_3=x_4=0.3\,, x_5=x_6=0.8, y=0.1\,, b=0.1\,$

$m^s = \{79.4, 121.5, 296.9, 354.3, 794.6, 808.8\}$ GeV\,;

$m^p = \{114.8, 296.9, 353.7, 794.6, 808.8\}$ GeV\,.
\end{itemize}

\begin{figure}[t]
  \begin{center} \hspace*{0mm}
      \psfig{file=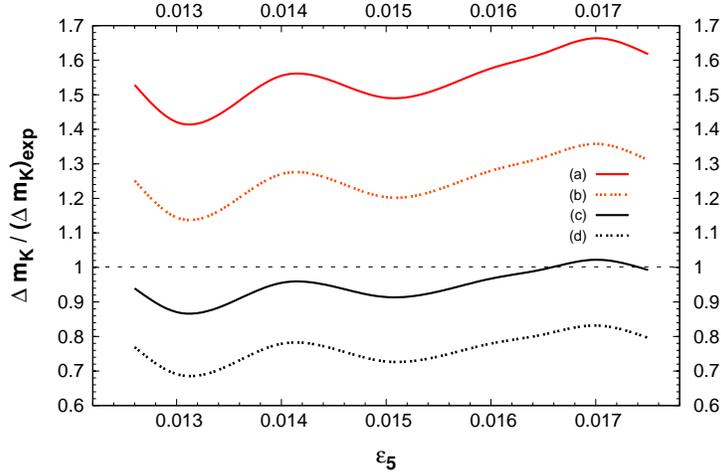,width=70mm,angle=270,clip=} 
    \caption{$\Delta m_K / (\Delta m_K)_{\text{exp}}$ as a function of 
$\varepsilon_5$ for set B and $\tan \beta=5$. The Higgs parameters
      correspond to textures (a)-(d).}
    \label{fig:dmk}
  \end{center}
\end{figure}

In Fig.~\ref{fig:dmk} we plot the ratio 
$\Delta m_K/(\Delta m_K)_{\text{exp}}$ versus
$\varepsilon_5$, for cases (a)-(d). We considered $\tan \beta=5$ and, 
since the other sets of input quark masses were in general associated
to smaller FCNC effects, we take the quarks masses 
as in ``set B'' (Table~\ref{set:ae}).
Once again, all the points displayed comply with the bounds from the 
CKM matrix.
>From Fig.~\ref{fig:dmk} it is clear that it is quite easy for the
orbifold model to accommodate the current experimental values for 
$\Delta m_K$. Even though the model presents the possibility of
important tree-level contributions to the kaon mass difference, 
all the textures considered give rise to contributions very close to the
experimental value. Although (a) and (b) are not in agreement with the
measured value of $\Delta m_K$, their contribution is within order of
magnitude of $(\Delta m_K)_{\text{exp}}$.
As seen from Fig.~\ref{fig:dmk}, with a
considerably light Higgs spectrum (i.e. $m_{h^0_i} < 1$ TeV), one is
safely below the experimental bound, as exhibited by cases (c) and (d). 
This is not entirely unexpected given the strongly hierarchical structure of
the Yukawa couplings (notice from Eq.~(\ref{Yd:1:5}) that $Y^d_{21}$
is suppressed by $\varepsilon_5^2$). 

One important aspect clearly manifest in Fig.~\ref{fig:dmk},
and which has been overlooked in some previous analyses, is that 
the Higgs mixings can be as relevant as the Higgs eigenvalues in 
determining the contributions to $\Delta m_K$.
This is patent in the comparison of curves (c) and
(d). From a na\"{\i}ve inspection of the Higgs spectra associated to 
each case, one would expect that (c) would induce a much
stronger suppression to the model's contribution to $\Delta
m_K$. Nevertheless, case (d), with a spectrum quite similar to case
(b), and indeed much lighter than that of (d), but with much smaller
mixings, is the one associated with the strongest suppression of 
$\Delta m_K$.
\begin{figure}[t]
  \begin{center} \hspace*{0mm}
      \psfig{file=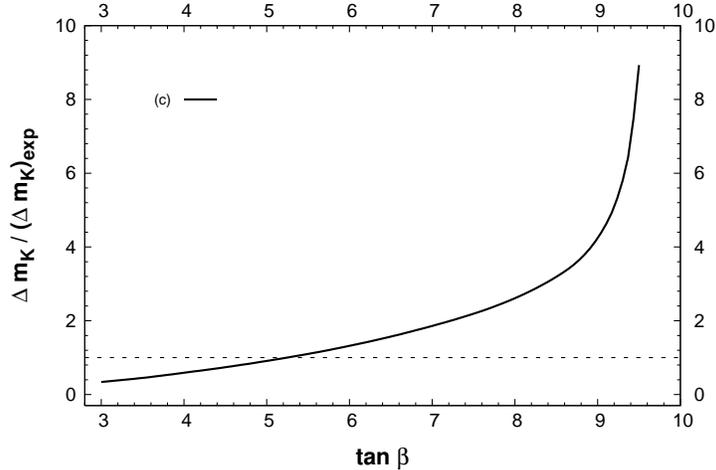,width=70mm,angle=270,clip=} 
    \caption{$\Delta m_K / (\Delta m_K)_{\text{exp}}$ as a function of 
      $\tan \beta$ for set B. The Higgs parameters
      correspond to texture (c).}
    \label{fig:dmk:tb}
  \end{center}
\end{figure}
It is also important to comment on the effect of changing $\tan \beta$
regarding the contributions to the kaon mass difference.
For the specific case of texture (c), let us investigate the effect of
varying $\tan \beta$. We take the quark masses as in set B, and while
keeping the Higgs parameters fixed, $\tan \beta$ is taken in the range
$3 \lesssim \tan \beta \lesssim 9.5$. 
As it becomes clear from Fig.~\ref{fig:dmk:tb}, larger
values of $\tan \beta$ produce increasingly larger contributions to
$\Delta m_K$. This is a direct consequence of the fact that, due to
larger off-diagonal terms in the Higgs mass matrices, the mixing is
larger, and the corresponding eigenstates become lighter. Even though
the masses of the heaviest states remain stable, the intermediate
states become lighter, and the FCNC contributions are less
suppressed. Close to $\tan \beta$=10, it is no longer possible to find
physical minima of the Higgs potential, and tachyonic states
emerge. This effect has been already pointed out in the general
analysis of~\cite{Escudero:2005hk}.

\subsubsection{\large $\pmb{B_d}$ and $\pmb{B_s}$ meson mass
  difference}
\begin{figure}[t]
  \begin{center} 
    \begin{tabular}{c}
      \psfig{file=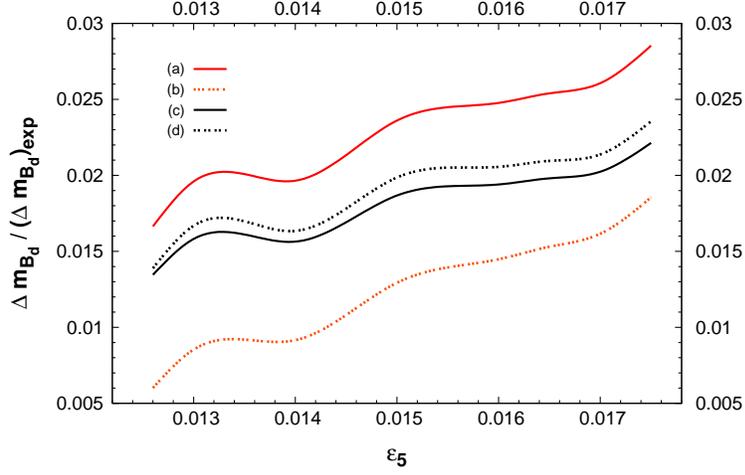,width=70mm,angle=270,clip=} 
      \end{tabular}
      \caption{$\Delta m_{B_d} / (\Delta m_{B_d})_{\text{exp}}$ 
as a function of $\varepsilon_5$ for set B and $\tan \beta=5$. 
The Higgs parameters correspond to textures (a)-(d).}
    \label{fig:dmbd}
  \end{center}
\end{figure}

For the several parameterisations of the Higgs sector used for the
analysis of $\Delta m_K$, we display in Fig.~\ref{fig:dmbd} the 
contributions of Higgs textures (a)-(d) to the ${B_d}$ mass difference.
As one would expect, given the structure of the mass matrices (and
thus the Yukawa couplings), the present model generates very small
contributions to $\Delta m_{B_d}$. All the textures analysed, even
those associated with excessive contributions to $\Delta m_K$ are in
good agreement with the experimental data on the $B_d$ mass difference. 
Notice that the behaviour
of the textures is now quite distinct: as an example, texture (b),
which generated the second largest contribution to the $\Delta m_K$ is
now the one associated with the strongest suppression. This stems from the
fact that the leading contributions are now given by 
distinct Higgses, whose couplings to the quarks need not be identical.

\begin{figure}[t]
  \begin{center} 
    \begin{tabular}{c}
      \psfig{file=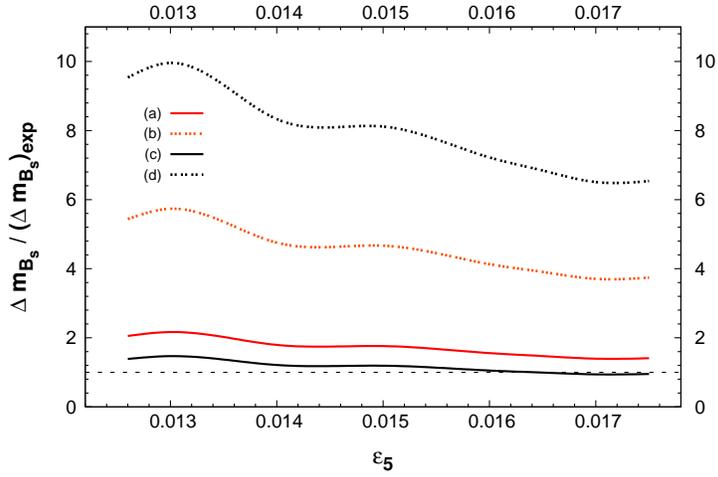,width=70mm,angle=270,clip=} 
    \end{tabular}
     \caption{$\Delta m_{B_s} / (\Delta m_{B_s})_{\text{exp}}$ 
as a function of $\varepsilon_5$ for set B and $\tan \beta=5$. 
The Higgs parameters correspond to textures (a)-(d).}
    \label{fig:dmbs}
  \end{center}
\end{figure}

Likewise, in Fig.~\ref{fig:dmbs} we plot the contributions to the 
${B_s}$ mass difference. In this case 
experiment only provides a lower bound, so that 
any ratio larger than 1 is in agreement with current data.
As we would expect from the discussion of Section~\ref{yukint},
this model's contributions to $\Delta m_{B_s}$ are well above the
current limit. 

\subsubsection{\large $\pmb{D^0-\bar D^0}$ mass difference}
The cases (a)-(d) considered in the previous subsections generate 
contributions to $\Delta m_D$ that exceed the experimental bounds by
at least a factor 10. As discussed in Section~\ref{yukint},
this is not surprising, nor excessively worrying. Nevertheless, and
for the sake of completing the analysis, let us consider four
additional Higgs patterns, in order to derive a bound on the mass of
the heaviest Higgs boson that would render the model compatible with
the data from the $D^0$ sector.

\begin{figure}[t]
  \begin{center} 
    \begin{tabular}{c}
      \psfig{file=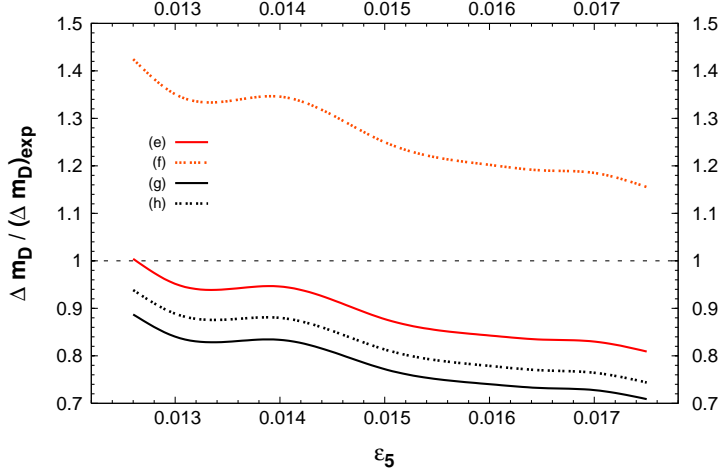,width=70mm,angle=270,clip=}
    \end{tabular}
     \caption{$\Delta m_{D} / (\Delta m_{D})_{\text{exp}}$ 
as a function of $\varepsilon_5$ for set B and $\tan \beta=5$. 
The Higgs parameters correspond to textures (e)-(h).}
    \label{fig:dmd}
  \end{center}
\end{figure}

The new Higgs textures are defined as follows:
\begin{itemize}
\item[(e)] 
$x_3=x_4=0.75\,, x_5=2.5\,, x_6=7.5\,, y=0.5\,, b=0.5\,$

$m^s = \{84.2, 672.7, 704.9, 1414, 2573, 7501\}$ GeV\,;

$m^p = \{673.1, 704.9, 1414, 2573, 7501\}$ GeV\,.

\item[(f)] 
$x_3=x_4=0.5\,, x_5=5\,, x_6=7.5\,, y=0.5\,, b=0.1\,$

$m^s = \{82.7, 201.4, 492.4, 516.4, 5000, 7500\}$ GeV\,;

$m^p = \{197.9, 492.4, 516.4, 5000, 7500\}$ GeV\,.

\item[(g)] 
$x_3=x_4=1\,, x_5=x_6=7.5\,, y=0.5\,, b=0.5\,$

$m^s = \{82.3, 378.5, 958.9, 1578, 7483, 7518\}$ GeV\,;

$m^p = \{379.2, 958.9, 1578, 7484, 7518\}$ GeV\,.

\item[(h)] 
$x_3=x_4=3\,, x_5=x_6=7\,, y=1\,, b=0.5\,$

$m^s = \{84.1, 1032, 2964, 3059, 6984, 7022\}$ GeV\,;

$m^p = \{1031, 2964, 3059, 6984, 7022\}$ GeV\,.
\end{itemize}
For the cases (e) to (h) we present in Fig.~\ref{fig:dmd} the
contributions to $\Delta m_D$. 
Compatibility with experiment can be obtained for any of the sets
(e), (g) or (h), thus suggesting that one of the Higgses (a state mostly
dominated by an up-type Higgs field) must be at least of around
7.5~TeV. Notice that no major hierarchy is required from the Higgs
spectrum - case (e)
is an excellent example of the latter statement, 
in the sense that one obtains states softly
ranging from 600 to 7500 GeV. 
One may wonder if such a choice of Higgs soft-breaking terms may lead
to a fine-tuning problem. In~\cite{Escudero:2005hk}, it was pointed
out that for non-degenerate VEVs, 
soft-breaking terms above the few TeV range typically 
induced a fine tuning stronger than 1\%.
Nevertheless, we again
stress that these values for the Higgs masses (as derived from the
$D^0$ mass difference analysis) should not be interpreted
from a very strict point of view. 

\subsection{Tree-level CP violation: $\pmb{\varepsilon_K}$}
Finally, we turn our attention to the issue of CP violation. So far,
we have seen that accommodating the several $\Delta F=2$ observables
is not an excessively hard task (especially if we choose to set aside
the $D^0$ sector). Nevertheless, a successful model of particle
physics must also comply with the observed CP violation in the kaon
sector. As we mentioned in Section~\ref{yukint}, we will only consider
the specific contribution of the present model to indirect CP
violation in the kaon sector ($\varepsilon_K$).
Therefore, we now assume $\varepsilon_5$ (and thus $\alpha^f$) to
be a complex quantity, and parameterise it as 
$\varepsilon_5 = |\varepsilon_5| \, e^{i \phi}$. 

\begin{figure}[t]
  \begin{center} 
    \begin{tabular}{c}
      \psfig{file=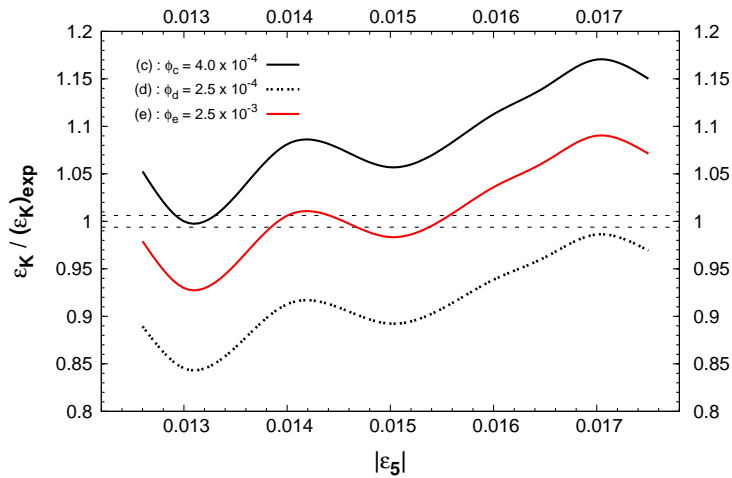,width=70mm,angle=270,clip=} 
    \end{tabular}
     \caption{$\varepsilon_K/ (\varepsilon_K)_{\text{exp}}$ 
as a function of $|\varepsilon_5|$ for set B and $\tan \beta=5$. 
The Higgs parameters correspond to textures (c)-(e).}
    \label{fig:ek}
  \end{center}
\end{figure}

In Fig.~\ref{fig:ek} we present the tree-level contributions to 
$\varepsilon_K$ (normalised by its experimental value) as a function
of $|\varepsilon_5|$. We take the input quark masses as in set B, fix
$\tan \beta=5$, and analyse the Higgs textures associated with cases
(c), (d) and (e). For each texture, the phase is assumed to be
$\phi_c=4.0 \times 10^{-4}$, $\phi_d=2.5 \times 10^{-4}$ and 
$\phi_e=2.5 \times 10^{-3}$.
These phases are taken as illustrative examples; we choose values that
for a specific set of input quark masses (set B, in this case) and a
given Higgs texture (c)-(e) 
simultaneously succeed in generating an amount of $\varepsilon_K$
close to the value experimentally measured (range delimited by 
dotted grey lines in Fig.~\ref{fig:ek}), and 
still have a compatible CKM. One should bare in mind that once 
$\arg \varepsilon_5$ (and thus $\arg \alpha^f$) is no longer a small
number, it will significantly affect the computation of the Yukawa
couplings, and thus the CKM matrix.
We have considered one texture that accommodates all FCNC observables,
namely texture (e), which already has a somewhat heavy Higgs
spectrum. In this case, the phase required to saturate 
$(\varepsilon_K)_{\text{exp}}$ is $\mathcal{O}(10^{-3})$.
Since we do not wish to view the $D^0$ sector as a very stringent
constraint, we also consider two other Higgs patterns, (c) and (d),
which only succeed in complying with both kaon and $B$-meson data. In these
cases, the phases necessary to obtain
$(\varepsilon_K)_{\text{exp}}$ 
are now $\mathcal{O}(10^{-4})$, as one would expect, since a
lighter set of Higgs bosons typically enhances the contributions.

In the range of parameters analysed in this plot, the amount of CP
violation stemming from the CKM matrix is $J_{CP} \sim
\mathcal{O}(10^{-8} - 10^{-6})$, i.e. at least one order of magnitude
below the SM value that is associated with the observed
$\varepsilon_K$~\cite{pdg2004}. 
The possibility of obtaining an orbifold configuration that saturates
the observed value of $\varepsilon_K$ and at the same time allows to
reproduce the $J_{CP}$ required by the unitarity triangle fits
should not be discarded. It is clear that the phase of $\varepsilon_5$
must be quite large, and~such values would push us to distinct areas
of the orbifold parameter space. It is worth emphasising that there
are still other sources of CP violation in addition to the one we have
studied in this Section. As mentioned in footnote~\ref{CPfoot}, a more
complicated choice of the VEVs $c_i$ could lead to physical phases in
the quark masses matrices, which would in turn contribute to the
physical CKM phase.

\section{Conclusions}\label{conc}
In this work we have investigated whether it is possible to find
abelian $Z_3$ orbifold configurations that are associated with
an experimentally viable low-energy scenario.

This class of models provides a beautiful geometric mechanism for the
generation of the fermion mass hierarchy. The Yukawa couplings are
explicitly calculable, and thus a solution to the
elusive flavour problem of the SM and MSSM can in principle be obtained.

We have concentrated here in $Z_3$ orbifold compactifications with two
Wilson lines, which naturally include three families for fermions
and Higgses. The fact that additional Yukawas are present opens the
possibility of obtaining realistic fermion masses and mixings,
entirely at the renormalisable level (with a key role being played by
the FI breaking).
We have surveyed the parameter space generated by the free orbifold
parameters, and we have verified that one can find configurations that
obey the EW symmetry breaking conditions, and can account for the
correct quark masses and mixings. 

On the other hand, 
having six Higgs doublets (and thus six quark Yukawa couplings)
poses the potential problem of having tree-level FCNCs. 
By assuming simple textures for the Higgs free parameters, we have
verified that the experimental data on the neutral kaon mass
difference, as well as on $\Delta m_{B_d}$ and $\Delta m_{B_s}$ can be
easily accommodated for a quite light Higgs spectra, namely 
$m_{h^0_i} \lesssim 1$~TeV. 
The data from the $D^0$ sector proves to be a more difficult
challenge, requiring a Higgs spectrum of at least 7 TeV, but we again
stress that, in view of the theoretical and experimental uncertainties
associated with the $D^0$ sector, this constraint should not be
over-emphasised.  

CP violation can be also embedded into the
low-energy theory. Although CP is a gauge symmetry of the full theory, 
it can be spontaneously broken at the string scale, if the VEVs of the
moduli have a non-vanishing phase. We have parameterised these effects
by assuming the presence of a phase in $\varepsilon_5$, and we have
verified that one can also obtain a value for $\varepsilon_K$ in
agreement with current experimental data.

The presence of a fairly light Higgs spectrum, composed by a total of 21
physical states, may provide abundant experimental signatures at
future colliders, like the Tevatron or the LHC. In fact, flavour
violating decays of the form $h_i \to q \bar q$, or $h_i \to l^+ l^-$
may provide the first clear evidence of this class of models.
$Z_3$ orbifold compactifications with two
Wilson lines are equally predictive regarding the lepton sector. This
analysis, especially that of the neutrino sector, will be addressed in
a forthcoming work~\cite{EJMT}.

\section*{Acknowledgements}
The work of N.~Escudero was supported by the ``Consejer\'{\i}a de Educaci\'on
de la Comunidad de Madrid - FPI program'', and ``European Social
Fund''. C.~Mu\~noz acknowledges
the support of the Spanish D.G.I. of the M.E.C. under ``Proyectos
Nacionales'' BFM2003-01266 and FPA2003-04597, and of the European
Union under the RTN program MRTN-CT-2004-503369. The work of A.~M.~Teixeira
is supported by ``Funda\c c\~ao para a Ci\^encia e Tecnologia'' under
the grant SFRH/BPD/11509/2002, and also by  ``Proyectos Nacionales'' 
BFM2003-01266.
The authors are all indebted to KAIST for the hospitality extended to
them during the final stages of this work, and also acknowledge the
financial support of the KAIST Visitor Program.

\end{document}